\documentclass[twocolumn,prd,aps,prl,floats,nofootinbib,floatfix,amsmath,superscriptaddress,
amssymb,longbibliography,secnumarabic]{revtex4-1} %

\usepackage{graphicx}
\graphicspath{{./figs/}}
\usepackage{dcolumn}
\usepackage{bm}
\usepackage{hyperref}
\usepackage{braket}
\usepackage{aas_macros} 

\usepackage{color}
\definecolor{mulberry}{rgb}{0.5,0,0.5}
\definecolor{orange}{rgb}{1,0.65,0}

\newcommand{\me}{\mathrm{e}}
\newcommand{\diff}{\mathrm{d}}

\begin{document}

\hfill 
\vspace{15pt}

\title{Dark Astronomy with Dark Matter Detectors}

\author{Gonzalo Alonso-\'Alvarez}
\email{gonzalo.alonso@utoronto.ca}
\author{David Curtin}
\email{dcurtin@physics.utoronto.ca}

\vspace*{3mm}

\affiliation{Department of Physics, University of Toronto, Toronto, ON M5S 1A7, Canada\\
}

\begin{abstract}
We present a novel way of probing non-gravitational dark matter interactions: \textit{dark astronomy}, which leverages the dark radiation emitted by dissipative dark sectors.
If the mediator of the dark matter self interactions is a dark photon with a small mass that kinetically mixes with the visible photon, the dark radiation flux becomes accessible to underground experiments.
We argue that the emission may be dominantly longitudinally polarized, thereby enhancing the sensitivity of direct detection experiments such as XENON and SENSEI to this signal.
We introduce a new detection mechanism based on resonant dark-photon-to-photon conversion at the surface of conducting materials, which offers unique directional sensitivity to dark radiation.
This mechanism facilitates the development of experiments that combine dark matter detection techniques with methods of traditional astronomy, opening the possibility to map dark radiation sources within our galaxy.
\end{abstract}

\maketitle

\section{Introduction}
\label{sec:introduction}

Beyond gravitational processes, dark matter (DM) may be looked for through direct interactions with visible matter. 
Currently, there exist two\footnote{In addition, DM can of course be produced at colliders, but verifying the DM nature of a new invisible state is more challenging.} main strategies to do so: direct detection, in which the DM particle interacts with a detector at Earth, and indirect detection, in which DM creates a flux of visible particles through e.g. decay or annihilation.

The direct detection landscape has boomed in the last few decades.
A strong push towards larger target volumes in noble gas detectors has led to strong constraints on the WIMP scenario~\cite{Akerib:2022ort}.
Recently, innovations in material science and detector technologies have lowered thresholds and background rates~\cite{Essig:2022dfa}, and will allow to probe thermal dark matter candidates with masses way below the weak scale~\cite{Knapen:2017xzo}.
In parallel, a rapidly growing experimental program~\cite{Antel:2023hkf} is targeting the typically sub-eV non-thermal dark matter frontier that includes particles like axions and dark photons.
Although prospects in these areas are promising, no positive detection has to date been reported.

Indirect detection has similarly experienced major developments~\cite{Leane:2020liq}, including expanding searches to new ranges of the electromagnetic and cosmic ray spectrum and using new messengers like gravitational waves~\cite{Bertone:2019irm} and neutrinos~\cite{Arguelles:2019ouk}.
Foregrounds from unknown or mismodeled baryonic processes make it difficult to claim an unambiguous detection, but several tantalizing hints currently exist~\cite{PAMELA:2008gwm,Hooper:2010mq,Siegert:2015knp,Cui:2016ppb,Cuoco:2016eej,AMS_AntiHe}.
Future missions and observatories, along with improved astrophysical modeling, should be able to shed light on these potential signals.

In this work, we propose a new avenue to probe the nature of dark matter: \textit{dark astronomy}, based on detecting the flux of dark radiation that the DM may emit if it has dissipative self-interactions. 
Such emission would be completely analogous to that of visible electromagnetic radiation due to astrophysical processes involving baryonic matter, which is our main means of gathering information about the visible universe.
This new paradigm blends  techniques of indirect an direct detection, since the target is a radiative signal generated by the galactic DM particles, but a dark one which must be detected in a shielded, low-background environment.

Dark matter self interactions have long been a subject of study due to their potential impact on cosmology~\cite{Cyr-Racine:2015ihg}, galaxy and cluster morphology~\cite{Tulin:2017ara}, and dark matter production~\cite{Bernal:2015ova}.
A possibility that is gathering increasing interest is that these interactions are inelastic and thus dissipative, like those that allow baryons to cool and form galactic structure.
These interactions arise naturally in dark sector scenarios (e.g.~\cite{Ackerman:2008kmp}), which predict the existence of multiple particles and forces between them.
Indeed, a $\gtrsim \mathcal{O}(10\%)$ mass fraction of DM could consist of a strongly dissipative component without being in conflict with current self-interaction bounds~\cite{Randall:2008ppe}.

A wealth of dissipative DM models of varying degrees of complexity exist.
The simplest are excited dark matter models that feature two (or more) states with a small mass splitting~\cite{Tucker-Smith:2001myb,Arkani-Hamed:2008hhe,Schutz:2014nka,Blennow:2016gde,Zhang:2016dck}.
These lead do different phenomenology compared to elastic self-interacting DM~\cite{2019MNRAS.484.5437V,Alvarez:2019nwt,2021MNRAS.500.1531C,2023MNRAS.524..288O,Leonard:2024mqo}, and the dark matter would shine in the mediator of the self interaction if its mass is smaller than the mass splitting.
Strongly interacting dark sectors can lead to dissipation through bound state formation~\cite{Detmold:2014qqa,Wise:2014jva,Krnjaic:2014xza,Gresham:2018anj}, or through emission of glueballs or light mesons in confined theories~\cite{Alves:2009nf,Boddy:2014yra,Cline:2013zca,Alonso-Alvarez:2023rjq}.
Models containing a long range force akin to electromagnetism (EM) are very effective at cooling thanks to the possibility of radiating the massless (or very light) mediator~\cite{Kaplan:2009de,Fan:2013yva,Foot:2014uba}.
Minimal implementations include a single charged fermion~\cite{Chang:2018bgx}, and atomic dark matter~\cite{Kaplan:2009de} with two oppositely charged states.
Some theoretically motivated constructions, like twin Higgs~\cite{Chacko:2005pe}  and mirror world~\cite{Kolb:1985bf,Glashow:1985ud} models, feature a combination of the above mechanisms.

The cosmological and astrophysical dynamics of dissipative dark sectors rival in complexity those in the visible sector and are highly model dependent.
Numerical simulations have been performed for some scenarios~\cite{Essig:2018pzq,Huo:2019yhk,Shen:2021frv,Shen:2022opd,Roy:2023zar,Gemmell:2023trd,Roy:2024bcu}, and a fairly generic prediction of efficient cooling~\cite{2022ApJ...934..120R} is the formation of dark matter compact objects~\cite{Buckley:2017ttd,Chang:2018bgx,2022ApJ...939L..12G}.
The end-stages of such a collapse can include black holes~\cite{DAmico:2017lqj,Shandera:2018xkn,Choquette:2018lvq,Latif:2018kqv}, dark white dwarfs~\cite{Ryan:2022hku} and dark neutron stars~\cite{Hippert:2021fch}. However, before reaching this terminal stage, dark compact objects can have a `main sequence life' where they either simply radiate away their gravitational potential energy as heat on their respective Kelvin-Helmholtz timescale, which could be more than billions of years, or generate large amounts of radiative energy through internal processes like dark nuclear reactions~\cite{Curtin:2019ngc, Curtin:2019lhm}. The resulting mirror or dark stars~\cite{Mohapatra:1996yy,Foot:1999hm,Berezhiani:2003xm, Curtin:2019ngc, Curtin:2019lhm} would enormously boost the expected dark radiation luminosity in the form of dark starlight.

In this paper, we perform a first study of the detection prospects of the dark radiation signal arising in dissipative dark matter models.
To remain general, we parametrize the dark radiation flux arriving at earth simply in terms of a black-body temperature and a total luminosity of the galactic ensemble of sources.
We focus on the scenario in which the dark radiation is made up of light (but not massless) dark photons that kinetically mix with the visible photon~\cite{Holdom:1985ag}.
The presence of a dark photon mass is crucial to enable dark-photon-to-photon conversions, which make the former detectable at dark matter experiments, as has been studied in~\cite{An:2013yua,Hochberg:2016sqx,Bloch:2016sjj,An:2020bxd} in the context of solar dark photons and dark photon DM.
In addition to examining the sensitivity of existing detectors like SENSEI~\cite{SENSEI:2020dpa} and XENON1T~\cite{XENON:2017lvq} to the dark radiation signal, in this work we propose a new class of directional experiments based on a surface conversion effect that we characterize for the first time.

Kinetically mixed light dark photons have been studied as mediators between the dark and the visible sector~\cite{Essig:2011nj} and as dark matter candidates themselves~\cite{Arias:2012az}.
Here we do not assume them to make up any significant fraction of the dark matter, but even in this case the kinetic mixing is strongly constrained from the production of dark photons in laboratory experiments and in astrophysical bodies~\cite{Fabbrichesi:2020wbt}, most prominently the sun~\cite{Redondo:2015iea,An:2013yfc}.
As we will see, solar emission constitutes an important foreground for the searches of dark radiation originating from dissipative dark matter structures.
Through the kinetic mixing, dark sector particles become millicharged under electromagnetism, with consequences for their production and detection~\cite{Iles:2024zka}.

This paper is organized as follows. We start by discussing the basis freedom of kinetically mixed massive dark photons in Sec.~\ref{sec:dark_photons}, where we also characterize their propagation in a visible medium, paying special attention to longitudinal modes.
The flux of dark photons at Earth is modeled in Sec.~\ref{sec:sources} considering two main sources: the sun and the radiative emission from dissipative dark matter in the galaxy.
In Sec.~\ref{sec:detection}, we outline the detection prospects of the dark radiation signal by DM direct detection experiments, and by a new experimental proposal exploiting the novel surface resonant conversion process. 
We summarize and conclude in Sec.~\ref{sec:conclusions}.

\section{Massive dark photons}\label{sec:dark_photons}

Our basic assumption is that the dark matter is charged under a dark U(1) gauge group whose gauge boson acquires a small mass through the St\"uckelberg mechanism and kinetically mixes with the Standard Model (SM) photon.
This construction allows dark matter to have dissipative interactions, but their exact nature depends on the specific matter content of the theory.
We remain agnostic about those details, and focus on the dynamics of the dark radiation interacting with visible matter once it has been emitted by dark matter through some unspecified astrophysical process.

\subsection{Kinetic mixing and basis freedom}

The minimal particle physics model thus consists of a massive gauge boson $A_\mu'$ kinetically mixed with electromagnetism via a  kinetic mixing parameter $\epsilon \ll 1$,
 \begin{eqnarray}\label{eq:L_kinetic_mixing}
    \mathcal{L} &=& -\frac{1}{4} F_{\mu \nu} F^{\mu \nu}
     -\frac{1}{4} F'_{\mu \nu} {F'}^{\mu \nu}
     -\frac{\epsilon}{2} F_{\mu \nu} {F'}^{\mu \nu}
     \\
     \nonumber &&
      + \frac{1}{2}m^2 A_\mu' A^{\prime\mu} + e j^\mu_{\rm SM} A_\mu + e' j^\mu_{\rm DM} A'_\mu.
\end{eqnarray}
Here, $j_{\rm SM}$ is the SM EM current, while $j_{\rm DM}$ is the dark current containing states charged under the dark U(1), some of which are assumed to make up (a fraction) of the DM. 
The dark photon mass is denoted by $m$.

The photon kinetic terms can be made canonical with a field redefinition $(A, A') \to (A_1, A_2)$.
%
This field redefinition is famously not unique since the kinetic terms preserve an $O(2)$ symmetry, which is only broken by the dark photon mass term and the couplings to matter.
Any physical observable is of course basis independent, but often there is a particularly convenient choice.

One of those is the vacuum propagation basis where the mass terms are diagonal.
It can be reached from Eq.~\eqref{eq:L_kinetic_mixing} by defining the vacuum massless state $V = A - \epsilon A'$ and keeping the massive one $A'$ unchanged.
The transformation also involves a small rescaling of the dark electric charge, kinetic mixing parameter, and the massive photon field; we do not introduce new notation for the rescaled quantities to simplify the presentation. In this propagation basis, the Lagrangian is
\begin{eqnarray}
    \mathcal{L} &=& -\frac{1}{4} V_{\mu \nu} V^{\mu \nu}
     -\frac{1}{4} F'_{\mu \nu} F^{\prime\mu \nu} + \frac{1}{2}m^2 A'_\mu A^{\prime \mu}
     \nonumber \\
      &&
       + e j^\mu_{\rm SM} (V_\mu - \epsilon A'_\mu) + e' j^\mu_{\rm DM} A'_\mu.\label{eq:L_SM_millicharged}
\end{eqnarray}
In this basis, dark charges only interact with the massive photon, while visible matter couples to both gauge bosons.
For this reason, it is also sometimes called the SM milli-charged basis.

Another convenient choice is the dark milli-charged basis which involves the SM-interacting state $A$ and the sterile state $S = A' + \epsilon A$ which does not directly couple to SM currents.
Together with a small rescaling of the visible photon field, the electric charge and kinetic mixing (we again do not rename the rescaled quantities for convenience), the Lagrangian in Eq.~\eqref{eq:L_kinetic_mixing} transforms into
\begin{eqnarray}
    \mathcal{L} &=& -\frac{1}{4} F_{\mu \nu} F^{\mu \nu}
     -\frac{1}{4} S_{\mu \nu} S^{\mu \nu} + \frac{1}{2}m^2 (S_\mu-\epsilon A_\mu)^2
     \nonumber \\
      &&
       + e j^\mu_{\rm SM} A_\mu + e' j^\mu_{\rm DM} (S_\mu - \epsilon A_\mu).\label{eq:L_DM_millicharged}
\end{eqnarray}
Although $S$ and $A$ here are not propagation eigenstates, this basis simplifies the calculation of dark photon emission from the sun and detection at Earth-based experiments, as only $A$ experiences in-medium effects when a mixed state propagates through visible matter.

So far our discussion has assumed no additional states associated with the mechanism that generates the dark photon mass, as is the case in St\"uckelberg scenarios~\cite{Stueckelberg:1938hvi}.
However, if the dark photon acquires its mass through a dark Higgs mechanism, a charged scalar $s$ appears in the spectrum, with a similar mass to the dark photon unless the dark gauge coupling is extremely small~\cite{Cline:2024wja}.
Through kinetic mixing, the theory contains an $\epsilon s\gamma\gamma'$ vertex~\cite{An:2013yua} that allows for $\gamma' \rightarrow \gamma s$ decays.
This results in a visible photon flux from dark astrophysical environments that can directly be looked for in concrete models. We defer study of this dark-Higgs-enabled signal for a future analysis.

It is important to note that the situation is drastically different if the dark photon is massless.
In the absence of a mass term, any linear combination of states with diagonal kinetic terms is a propagation eigenstate.
Even if dark charges radiate a state that has a small admixture of visible photon, for detection purposes this visible component is absorbed by any shielding that is unavoidably needed to reduce foregrounds.
After shielding, the only way to enable dark photon-to-photon oscillations is via an effective dark photon mass, which can be generated by e.g. a local dark plasma like is evoked in Ref.~\cite{Berlin:2023gvx} for light-shining-though-the-wall experiments and in Ref.~\cite{Berlin:2022hmt} for cosmological observables.

\subsection{Dark photon propagation}
\label{sec:propagation}

In the absence of SM charges, the propagation basis in~\eqref{eq:L_SM_millicharged} gives a straightforward picture of the dark photon dynamics: $A'$ is both an interaction and propagation eigenstate.
Thus, dark charges emit dark photons with a well defined mass, and the massless field is never excited.
As we will see in the next section, the longitudinal component of a massive gauge boson is more weakly coupled than the transverse one.
This means that its emission from high-density environments can be favored compared to that of transverse modes, which get trapped when the opacity is high.

The dynamics are more involved when visible and/or dark photons propagate in media containing SM charges.
In this case, it is convenient to start with the DM millicharged basis of~\eqref{eq:L_DM_millicharged}, in which the sterile state $S$ does not feel the SM background.
This comes at the cost of introducing the mass mixing term that induces oscillations between $A$ and $S$.
In a SM medium, $A$ acquires an effective mass that changes the propagation eigenstates: a new material-dependent mass basis must be identified to construct the wavefunctions.

This procedure has been followed in different works to understand the production of dark photons in the sun~\cite{Redondo:2013lna,Redondo:2015iea} as well as the detection of transverse and longitudinal modes in helioscopes~\cite{Redondo:2008aa,OShea:2023gqn} and light-shining-through-the-wall experiments~\cite{Ahlers:2007rd,Graham:2014sha} (other studies choose to work in the basis in which kinetic mixing is explicitly present~\cite{An:2013yfc,An:2014twa}). 
In a visible plasma like that of the sun, dark photons are dominantly produced through resonant photon-to-dark-photon conversions.
For transverse modes, these occur in regions where the plasma frequency of the medium matches the mass of the dark photon, while longitudinal modes are produced at the local plasma frequency as long as their mass is comparatively small.
On the detection side, transverse modes are challenging because any shielding before the detector projects out the visible component of the incoming dark photon flux: the visible component is only regenerated through mass mixing, resulting in an extra $(m d)^2$ suppression, where $d$ is the conversion length between shielding and detector.
In contrast, longitudinal modes are not shielded in this way since the visible vacuum propagation eigenstate is massless and thus cannot support a longitudinal component.
As was pointed out in Ref.~\cite{Graham:2014sha}, this results in a parametric enhancement of the longitudinal dark photon detection prospects.

Comparably, the dynamics of longitudinal dark-photon-to-photon conversion in solid state materials has not been explored in much detail.
Here, we employ a two-state flavor mixing formalism to show that in addition to the volume absorption described in~\cite{An:2013yua}, there exists a previously overlooked \textit{surface resonant conversion} effect that can be exploited in a new class of experiments with directional sensitivity to the dark photon flux.
We also identify a nonresonant contribution to the surface conversion linked to the non-shieldable nature of longitudinal dark photons, which is not our current focus since it appears harder to exploit experimentally.

Longitudinal dark photon modes do not oscillate in vacuum, since the massless photon has no longitudinal component for them to mix with.
However, in a medium with free charges, the electromagnetic field acquires an effective third polarization due to collective charge oscillations, usually called plasmons (see Appendix~\ref{app:dispersion_relation} for a brief review of in-medium photon dispersion effects).
The longitudinal propagation eigenstates therefore become admixtures of dark photon and plasmon.

In the dark milli-charged basis of Eq.~\eqref{eq:L_DM_millicharged}, the equations of motion for the longitudinal modes of the kinetically mixed photon-dark photon system in a visible medium can be written as
\begin{equation}\label{eq:basis_rotation}
    \begin{pmatrix}
    \omega^2-k^2-\pi_L & -\epsilon m^2 \\
    -\epsilon m^2 & \omega^2 - k^2 - m^2
    \end{pmatrix}
    \begin{pmatrix}
    A \\ S
    \end{pmatrix} = 0.
\end{equation}
with the self-energy of the longitudinal photon as defined in Eq.~\eqref{eq:photon_self_energy}.
Let us focus on the case of a dark photon plane wave in the massive vacuum propagating eigenstate $A'$ in Eq.~\eqref{eq:L_SM_millicharged} entering a conductor medium from vacuum.
We model the conductor as a degenerate gas of electrons with a small Fermi velocity $v_F\ll 1$.
The dispersion relation of longitudinal plasmons for general $\omega$ and $k$ was recently calculated in~\cite{Scherer:2024uui}, recovering the Lindhard response function~\cite{Dressel_Grüner_2002} for nonrelativistic degenerate electrons.

We are dealing with modes near the light cone, since the incoming dark photon in vacuum satisfies the dispersion relation $\omega^2 = k^2 + m^2$ with $m\ll\omega$.
In that limit, to leading order in $v_F$, the self energy of the longitudinal mode is
\begin{equation}
    \pi_L = \frac{\omega^2 - k^2}{\omega^2} \pi_T.
\end{equation}
The real part of the transverse self energy is simply $\mathrm{Re}\,\pi_T = \omega_p^2$, where the plasma frequency is given by $\omega_p^2=4\alpha p_F^2 v_F /3\pi$, with the Fermi momentum $p_F=m_ev_F$~\cite{Raffelt:1996wa}.
The imaginary component describing the losses in the material also gets a contribution from purely electronic processes, but in solid state conductors the dominant source of losses are electron scatterings with ions in the lattice and impurities.
This contribution can be inferred from measurements of the conductivity as a function of frequency.

With this, the EOMs in Eq.~\eqref{eq:basis_rotation} can be rewritten as
\begin{equation}
    \begin{pmatrix}
    Z_L^{-1} (\omega^2-\pi_T) & -\epsilon m^2 \\
    -\epsilon m^2 & \omega^2 - k^2 - m^2
    \end{pmatrix}
    \begin{pmatrix}
    A \\ S
    \end{pmatrix} = 0.
\end{equation}
where we have introduced the renormalization factor $Z_L = \omega^2/(\omega^2-k^2)$.
These can be brought into canonical form by renormalizing the $A$ field via $A\rightarrow\sqrt{Z_L}A$, leading to
\begin{equation}
    \begin{pmatrix}
    \omega^2-\pi_T & -\tilde{\epsilon}m^2 \\
    -\tilde{\epsilon} m^2 & \omega^2 - k^2 - m^2
    \end{pmatrix}
    \begin{pmatrix}
    A \\ S
    \end{pmatrix} = 0.
\end{equation}
The canonical photon field thus mixes with renormalized strength $\tilde\epsilon = \sqrt{Z_L}\epsilon$ (its gauge coupling to external charges is also similarly renormalized).
The renormalization also suppresses the overlap between the in-medium state $A$ and the vacuum massive propagation eigenstate $S$.
To leading order in $\tilde\epsilon$, the system can be diagonalized to obtain the in-medium propagating states via the field redefinition
\begin{equation}
\label{eq:field_redef_diag}
    \begin{pmatrix}
    \tilde{A} \\ \tilde{S}
    \end{pmatrix} = 
    \begin{pmatrix}
    1 & -\theta_{\omega,k} \\
    \theta_{\omega,k} & 1
    \end{pmatrix}
    \begin{pmatrix}
    A \\ S
    \end{pmatrix},
\end{equation}
where we have defined the in-medium mixing angle
\begin{equation}\label{eq:mixing_angle}
    \theta_{\omega,k} = \frac{\tilde{\epsilon}m^2}{k^2 + m^2 - \pi_T(\omega,\,k)}.
\end{equation}
This leads to the equations of motion
\begin{align}
    (\omega^2 - \pi_T) \tilde{A} &= 0,\label{eq:EOM_Atilde}\\
    \left( \omega^2 - k^2 - m^2 + \tilde{\epsilon}m^2\theta_{\omega,k} \right) \tilde{S} &=0, \label{eq:EOM_Stilde}
\end{align}
where we have kept the $\mathcal{O}(\epsilon^2)$ term in~\eqref{eq:EOM_Stilde} as it gives the leading-order absorption of the (almost) sterile eigenstate.

The real part of the EOMs can be separately solved to obtain the dispersion relation of the two components.
For small kinetic mixings, Eq.~\eqref{eq:EOM_Stilde} simply leads to $\omega_S = \sqrt{k^2+m^2}$.
The dispersion relation for $\tilde{A}$ is to leading order that of a longitudinal plasmon, $\omega_A = \omega_p$.
The two propagation eigenstates are thus $\tilde{A}$, mostly plasmon-like, and $\tilde{S}$, mostly sterile but with a small active admixture that allows a visible medium to absorb it at an $\epsilon^2$-suppressed rate.

Equipped with the dispersion relations, we can linearize the EOMs to obtain Schr\"odinger-like equations for timelike modes near the light cone,
\begin{align}
    i\partial_t\tilde{A} &= \left( \frac{k}{2} + \frac{\omega_p^2}{2k} + \frac{i}{2k} \, \mathrm{Im}\,\pi_T \right) \tilde{A},\label{eq:linear_EOM_Atilde}\\
    i\partial_t\tilde{S} &= \left( \omega_S + \frac{i}{2k} \, \tilde{\epsilon}m^2 \mathrm{Im}\, \theta_{\omega_S,k} \right) \tilde{S}. \label{eq:linear_EOM_Stilde}
\end{align}

An incoming dark photon plane wave of frequency $\omega$ in the vacuum propagating eigenstate $A'$ contains an $1-\mathcal{O}(\epsilon^2)$ component of $\tilde{S}$ in the medium. 
To leading order in $\epsilon$, it is thus damped through this admixture at a rate 
\begin{align}\label{eq:absorption_rate_volume}
    \Gamma_{\rm vol} &= -\frac{\diff |\tilde{S}|^2}{\diff t} = \frac{\tilde{\epsilon} m^2}{\omega} \mathrm{Im}\,\theta_{\omega,k} \nonumber \\  
    &= \frac{\epsilon^2 m^2 \mathrm{Re}\,\sigma}{\omega^2 |n|^4},
\end{align}
which occurs throughout the whole volume of the material.
We have expressed the final result in terms of the complex index of refraction of the material $n^2 = 1 - \pi_T/\omega^2$ and the complex conductivity $\sigma = i \pi_T/\omega$.
These should be evaluated at $k=\sqrt{\omega^2-m^2}$, but for small dark photon masses $m\ll\omega$ it is a good approximation to evaluate them at the light cone $k=\omega$.
This agrees with Refs.~\cite{An:2013yua,Hochberg:2016sqx} and features the usual $m^2/\omega^2$ suppression of the rate for small dark photon masses.

The EOMs also show the existence of a mostly plasmon-like propagation eigenstate that gets damped at a faster rate, unsuppressed by the small kinetic mixing.
The dark photon entering the conductor from vacuum is necessarily in the vacuum massive eigenstate $A'$ from Eq.~\eqref{eq:L_SM_millicharged}, which can be decomposed as $S - \epsilon A$ in the basis of Eq.~\eqref{eq:L_DM_millicharged}. 
After the basis rotation in Eq.~\eqref{eq:basis_rotation}, we see that the incoming dark photon gives rise to a small excitation of the  in-medium visible-plasmon-like $\tilde{A}$ field, with initial amplitude $Z_L^{-1/2}\epsilon + \theta_{\omega,k}$.
This component gets quickly damped, resulting in a strongly depth-dependent absorption rate 
\begin{align}\label{eq:absorption_rate_surface}
    \Gamma_{\rm surf} &= -\frac{\diff |\tilde{A}|^2}{\diff t} = -|Z_L^{-1/2}\epsilon + \theta_{\omega,k}|^2 \, \frac{\mathrm{Im}\,\pi_T}{\omega} \, \me^{\frac{\mathrm{Im}\,\pi_T}{\omega} z},
\end{align}
where the photon self-energy is to be evaluated at $k=\sqrt{\omega^2-m^2}$, and $z$ is the distance to the interface.

The absorption rate in Eq.~\eqref{eq:absorption_rate_surface} has two qualitatively different components.
The \textit{nonresonant} contribution associated with the $\propto Z_L^{-1/2}\epsilon$ term is due to the fact that the vacuum propagation eigenstate contains an interacting component that gets absorbed upon entering a visible medium.
It has a smooth frequency response dictated by the conductivity of the material.
The \textit{resonant} contribution linked to the $\theta_{\omega,k}$ term is due to longitudinal dark photon mixing with on-shell plasmons in the medium, which as Eq.~\eqref{eq:mixing_angle} shows is resonantly enhanced when $\omega = \omega_p$ for modes near the light cone.
For this frequency, the absorption rate in Eq.~\eqref{eq:absorption_rate_surface} becomes maximal,
\begin{equation}\label{eq:absorption_rate_surface_resonance}
    \Gamma_{\rm surf}\Big\rvert_{\omega_p} = \frac{\epsilon^2m^2 }{\omega_p^2} \mathrm{Re}\,\sigma({\omega_p}) \left( 1 + \frac{\omega_p^2}{\left[\mathrm{Re}\,\sigma({\omega_p})\right]^2} \right) \me^{-z \, \mathrm{Re}\,\sigma({\omega_p})}.
\end{equation}
If $\mathrm{Re}\,\sigma(\omega_p)\ll \omega_p$, the nonresonant contribution is negligible compared to the resonant one around the plasma frequency.
In that case and for $\omega$ near $\omega_p$, the surface rate is close to the volume one at the interface, i.e.~$\Gamma_{\rm surf} \simeq \Gamma_{\rm vol}$ at $z=0$.

The absorption rate in~\eqref{eq:absorption_rate_surface} is a surface effect: it is localized within a few mean free paths from the material's interface.
It can be understood as a result of the difference between the propagation eigenbases when transitioning from vacuum, where only the massive longitudinal dark photon propagates, to the conductor, where the plasmon mode exists as well.
One may be surprised that these longitudinal plasmons have a mean free path in the material that is a short as that of transverse modes. This is due to the wavefunction renormalization within the medium, which also causes the enhancement of the production rate of longitudinal modes as explained in~\cite{An:2013yfc,Redondo:2013lna}.

\begin{figure}[t]
\centering
\centerline{\includegraphics[width=0.99\linewidth]{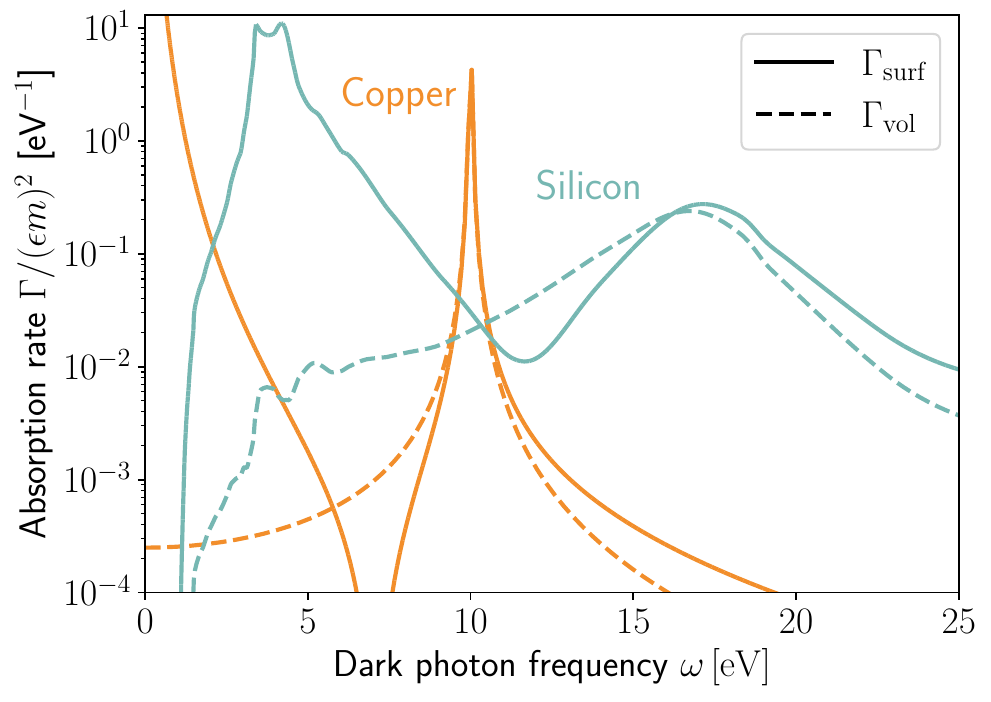}}
\caption{
Surface and volume longitudinal dark photon absorption rates for copper and silicon.
The surface rate is evaluated at the interface, i.e.~$z=0$, and we assume $m\ll10$~eV, see text for more details.}
\label{fig:GammaAS}
\end{figure}

The resonant dark-photon-to-plasmon conversion occurs because the in-medium mixing angle Eq.~\eqref{eq:mixing_angle} is maximized when the frequency of the incoming radiation matches the plasma frequency of the material.
If $\mathrm{Im}\,\pi_T \ll \mathrm{Re}\,\pi_T$, or equivalently $\mathrm{Re}\,\sigma(\omega_p) \ll \omega_p$, the resonance is narrow and the frequency-integrated absorption rate can be calculated using the narrow width approximation.
It is important to note that the resonant enhancement is independent of the value of the dark photon mass (as long as $m\ll\omega_p$), contrary to what occurs for transverse modes~\cite{OShea:2023gqn}.
As we will see, these properties make this process particularly promising as a way to detect massive dark photon radiation.

Figure~\ref{fig:GammaAS} compares the volumetric and surface absorption rates for copper and silicon.
Copper is treated as a Drude-Sommerfeld conductor with a plasma frequency $\omega_p=10$~eV.
For this figure, we fix $\mathrm{Re}\,\sigma(\omega_p)=\omega_p / 400$, representative of room temperature general-use-grade copper, though this quantity is highly temperature and purity dependent, as will be discussed.
For silicon, we infer the photon dispersion relations from the room-temperature values of the conductivity given in~\cite{Hochberg:2016sqx}.
Silicon only effectively behaves as a conductor for frequencies above its band gap energy of $\sim 1.2$~eV, and the resonance around its plasma frequency of $\sim 17$~eV is much broader than for copper.
This is linked to the photon mean free path at the resonance being much smaller in silicon than copper, which makes the resonant conversion process much more challenging to exploit in the semiconductor.

At frequencies away from the plasma frequency of the material, the surface absorption rate is dominated by the nonresonant contribution.
Although Fig.~\ref{fig:GammaAS} shows a steep rise of $\Gamma_{\rm surf}$ at low frequencies for both copper and silicon, this is due to a increase in the conductivity at those frequencies, which in turn concentrates the conversion in a smaller volume closer to the interface.
When accounting for this, the surface conversion at those frequencies is not enhanced compared to the volumetric absorption except right at the interface, which makes nonresonant conversion challenging to exploit experimentally as a way to absorb dark photons.

At the interface ($z=0$), volume and surface absorption have very similar rates, but $\Gamma_{\rm surf}$ quickly drops after a few absorption lengths.
The frequency response around the resonance is similar for both surface and volume processes, and thus their frequency-integrated rates are very similar within a mean free path of the interface.
If the dark photon mass is small compared to the frequencies of interest, as we assume here, it only enters as a multiplicative factor rescaling the amplitude of the absorption rates.

\section{Astrophysical sources of massive dark photons}\label{sec:sources}

\subsection{The Sun}
Any source of visible photons unavoidably produces a small amount of dark photons via their kinetic mixing, and so does the brightest photon source in our vicinity, the sun.
This constitutes an irreducible flux at earth irrespective of any assumption regarding the role of dark photons as force mediators in the dark sector.

The solar emission spectrum of massive dark photons has been studied in great detail in the literature.
Transverse mode emission was first characterized in~\cite{Redondo:2008aa} and detailed in~\cite{Redondo:2015iea}, while the radiation of longitudinal dark photons was first correctly described in~\cite{An:2013yfc} and further refined in~\cite{Redondo:2013lna}.

For small dark photon masses, longitudinal emission is dominant to transverse emission since the latter is relatively suppressed by a factor of $m^2/\omega^2$, where $\omega$ is the frequency of the radiation.
The relative enhancement of the longitudinal mode production rate can be traced back to its different in-medium dispersion relation (see Appendix~\ref{app:dispersion_relation}), which induces a renormalization of the kinetic mixing by a factor $\sqrt{Z_L}\sim \omega/m$ within a plasma like the solar one.
Thus, expect for small frequencies $\omega\lesssim m$, the longitudinal emission overwhelms the transverse one.

\begin{figure}[t]
\centering
\centerline{\includegraphics[width=0.95\linewidth]{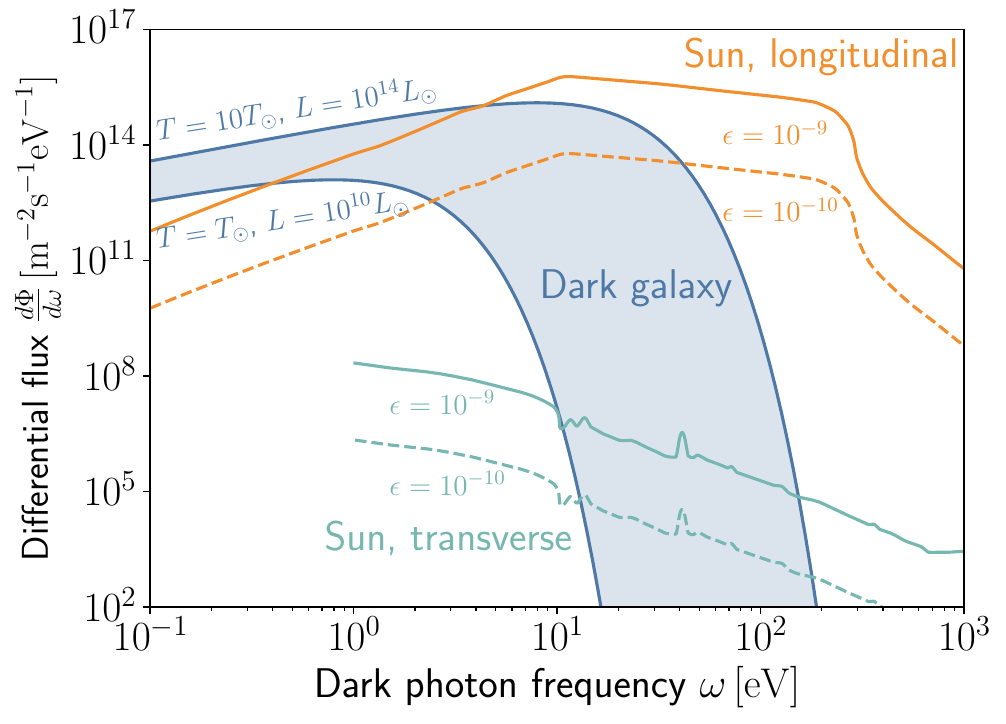}}
\caption{
Spectrum of dark photons at earth from astrophysical sources. 
Emission from the sun of longitudinal and transverse dark photons is shown in orange and green, for $m_{\gamma'}=1$~meV and  kinetic mixing parameters $\epsilon=10^{-9}$ (solid lines) and $\epsilon=10^{-10}$ (dashed lines). 
A hypothetical emission by the dark matter near the galactic center, characterized as a black body, is shown in blue.
The benchmark temperatures and luminosities represent a collection of dark stars making up a visible Milky-Way-like bulge.
}
\label{fig:flux_sources}
\end{figure}

Fig.~\ref{fig:flux_sources} displays the solar dark photon flux in both polarizations for an exemplary value of $m=1$~meV, which have been calculated following~\cite{Redondo:2015iea,An:2013yfc}.
For this small dark photon mass, the enhancement of the longitudinal mode compared to the transverse one is evident.
Note that we cut off the transverse spectrum below 1~eV, where the luminosity becomes very sensitive to the chemical composition in the outer layers of the sun and is challenging to model.
Such low frequencies are below the sensitivity of the experimental setups that we consider and thus irrelevant for the purposes of this paper.

\subsection{The Dark Galaxy}
If dark matter, or a subcomponent of it, is made up of states that can dissipate energy via dark photon emission, a whole spectrum of dark photons would arrive at Earth from different dark astrophysical sources, much like is the case for visible photons.
Compared to solar emission, emission of dark photons by dark sources is not $\epsilon^2$ suppressed, which can result in a much larger flux for small kinetic mixing parameters.

Characterizing this dark galactic emission in a concrete model is a formidable task, which is compounded by the fact that there exist many different constructions of dissipative dark sectors.
The emission may originate from a diffuse dark matter component, or be dominated by the luminosity of compact objects that may be the endpoint of the  cooling and collapse process in the presence of dissipative dark matter self interactions~\cite{Buckley:2017ttd,Chang:2018bgx,2022ApJ...939L..12G}.
A paradigmatic case is the formation of dark stars~\cite{Mohapatra:1996yy,Foot:1999hm,Berezhiani:2003xm, Curtin:2019ngc, Curtin:2019lhm}, whose luminosity could be powered by internal conversion processes like dark nuclear reactions,  simple cooling on a sufficiently-long Kelvin-Helmholtz time scale, or other processes that may turn on at high densities and pressures.

To avoid this complication and to remain general, we opt for a simple approach. We characterize the total dark galactic emission by a thermal spectrum at temperature $T$ and total luminosity $L$.
Furthermore, we assume that the sources that make up the bulk of the signal are concentrated around the galactic center, and are thus located $d\simeq 8$~kpc away from earth.
Thus, the dark photon number flux at Earth can be expressed as
\begin{equation}\label{eq:flux_dark_galaxy}
    \frac{\diff\Phi}{\diff\omega} = \frac{L}{4\pi d^2} \frac{15}{\pi^4T^4} \frac{\omega^2}{\me^{\omega/T}-1}.
\end{equation}
This flux is compared to the dark photon emission from the sun in Fig.~\ref{fig:flux_sources} for a range of representative values motivated by a hypothetical collection of dark stars located near the center of the galaxy.
The benchmark curves correspond to a visible-Milky-Way-like dark galaxy and one with 10 times hotter stars (and thus $10^4$ times larger luminosity as per the Stefan-Boltzmann law).
Since the dark galactic flux is independent of the kinetic mixing parameter while the solar one is proportional to $\epsilon^2$, the benchmark galaxies chosen outshine the sun for small kinetic mixings that will be accessible to upcoming detectors, as we argue in the next section.
Of course, other scenarios could lead to very different values for $L$ and $T$.

This very simplistic parametrization has the advantage of being easily applicable to other dark astrophysical sources, like an individual star located in the local environment of the solar system.
For an experiment with a limited field of view, one would need to rescale the flux by the relative angular size of its field compared to the angular extent of the source.
As reference, the bulge of the Milky Way extends to approximately $10^\circ$ in the sky, and galactic simulations of atomic dark matter~\cite{Roy:2023zar} show that dissipation in the dark matter can similarly lead to a concentration of dark objects in a compact bulge near the galactic center.

For massive dark photons, detection of longitudinal modes is parametrically favored compared to transverse ones.
Thus, a crucial quantity to predict is the relative fraction of longitudinally to transversely polarized massive dark photons that arrive at the detector.
Same as the total flux, this is very model dependent.
Naively, one could expect the longitudinal emission to be suppressed by $(m/\omega)^2$ compared to the transverse one due to the weaker coupling of the former.
However, this is not necessarily the case for all radiative processes and, most importantly, it does not take into account in-medium effects that are important in dense DM environments.

As a toy example, in Appendix~\ref{app:sun_L_emission} we calculate the emission of longitudinally polarized light from a sun-like star, i.e.~how the sun would shine if the visible photon were to suddenly acquire a small mass.
Fig.~\ref{fig:L_emission_sun} shows the spectral luminosity for different values of the photon mass.
In addition to the low-frequency emission from the photosphere, the presence of a small mass allows for volumetric emission from the stellar interior that can overwhelm the luminosity by many orders of magnitude.
The reason for this is easy to understand: the interaction rate of longitudinal modes of frequency $\omega$ are suppressed by $m^2/\omega^2$ compared to transverse ones.
For small masses, longitudinal modes become weakly coupled and can escape from the solar interior, enormously enhancing the luminosity at frequencies close to the peak of the thermal distribution at the temperature of the solar core.
This is similar to what occurs for weakly coupled particles such as axions or massive kinetically mixed dark photons in visible stars~\cite{Raffelt:1996wa}.

This of course cannot describe a realistic scenario, since such huge luminosities would very quickly backreact on the solar model that has been kept static in our calculation.
One could imagine that the outer layers of the star would collapse down to the photosphere of the longitudinal photon, generating a strong temperature and density gradient.
This would effectively bring the transverse photosphere close to the longitudinal one and balance the emission differential.
Although a detailed dark stellar evolution calculation is beyond the scope of this work, this picture motivates considering longitudinal mode emission from dark stars to be \emph{at least} comparable to transverse emission.

Of course, other dissipative dark matter scenarios can lead to different longitudinal-to-transverse photon emission ratios.
In what follows, we simply assume that the luminosity of the dark galaxy is predominantly longitudinally polarized.
For dark sector models where this is not necessarily the case, the limits apply to the longitudinal component of their dark radiation flux.

\section{Detection of dark photons at earth}\label{sec:detection}

\subsection{Dark matter experiments}

Dark photons can be absorbed in the target material of a DM direct detection experiment.
The absorption rate is given by Eq.~\eqref{eq:absorption_rate_volume} for longitudinal modes, which is enhanced by a factor of $\omega^2/m^2$ compared to that of transverse modes~\cite{An:2013yua} and thus dominates the signal rate for small dark photon masses.
Assuming that the surface effect can be neglected compared to the volume one, as is the case for experiments with large targets, the ensuing spectral event rate can be written as
\begin{equation}\label{eq:event_rate_volume}
    \frac{\diff R_{\rm vol}}{\diff E} = \frac{1}{\rho\,\beta} \, \frac{\diff\Phi}{\diff\omega}  \, \Gamma_{\rm vol},
\end{equation}
where $\omega=E$ for total absorption processes, $\rho$ is the density of the target material, and $\beta = \sqrt{1-(m/E)^2}$.
Searches for solar dark photons have been performed at various DM experiments, and we derive for the first time the limits on the dark radiation signal below.

Longitudinal solar dark photon emission is maximal at $10-100$~eV energies that are accessible to large-volume gas and liquid detectors.
The first limits were derived in~\cite{An:2013yua} using XENON10 data, and were shown to improve upon solar cooling bounds~\cite{Gondolo:2008dd}.
Limits were later derived for XENON100 and CDMSlite in~\cite{Bloch:2016sjj}.
Currently, the strongest constraints on solar dark photons come from the recast~\cite{An:2020bxd} of a S2-only analysis~\cite{XENON:2019gfn} of XENON1T data.
A later search by the XENON1T collaboration with sensitivity down to a single photoelectron~\cite{XENON:2021qze} found slightly weaker limits due to its smaller exposure.

\begin{figure*}[t]
\centering
\centerline{\includegraphics[width=0.65\linewidth]{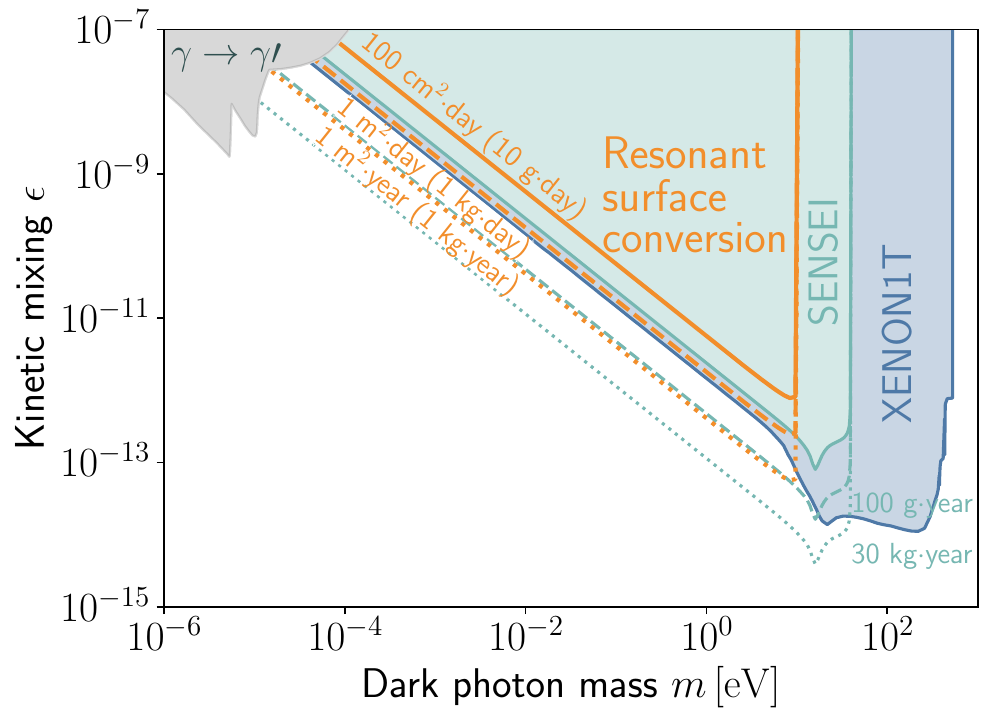}}
\caption{
90\% C.L.~limits on solar dark photon emission from the SENSEI (green) and XENON1T (blue) DM direct detection experiments, as well as projections for an experiment exploiting the novel surface conversion process (orange).
The assumed exposures for each line are indicated in the figure.
The region shaded in grey is excluded by a combination of light-shining-through-the-wall experiments~\cite{Betz:2013dza,Romanenko:2023irv}, CMB observables~\cite{McDermott:2019lch,Caputo:2020bdy,Chluba:2024wui,Arsenadze:2024ywr}, and tests of the Coulomb law~\cite{Jaeckel:2010xx,Kroff:2020zhp}.
}
\label{fig:sun}
\end{figure*}

Solid state DM detectors typically feature lower thresholds but a smaller exposure, which hinders their performance for the relatively high-energy solar dark photon signal.
As an example, in Fig.~\ref{fig:sun} we show limits from the SENSEI experiment~\cite{SENSEI:2023zdf,SENSEI:2024yyt} that we recast following the procedure described in Appendix~\ref{app:SENSEI}.
We also show projections for the planned SENSEI exposure of 100~g$\cdot$year~\cite{SENSEI:2020dpa} and for a 30~kg$\cdot$year experiment~\cite{2022arXiv220210518A}, assuming current backgrounds to scale proportional to exposure.

Direct detection experiments are also sensitive to the dark starlight signal.
We use the flux from Eq.~\eqref{eq:flux_dark_galaxy} to predict the volumetric absorption rate in silicon and xenon via Eq.~\eqref{eq:event_rate_volume} and calculate the sensitivity to the longitudinal luminosity of the dark galaxy from SENSEI and XENON1T following Appendix~\ref{app:SENSEI} and~\ref{app:XENON1T}, respectively.
The luminosities that can be accessible to these experiments as a function of black-body emission temperature correspond to the shaded regions in Fig.~\ref{fig:dark_galaxy}.

The luminosity in Fig.~\ref{fig:dark_galaxy} is normalized assuming that the dark photon source is located at the galactic center, 8 kpc away, but it can easily be rescaled to other distances.
The solid lines represent the smallest possible luminosity that the experiments can be sensitive to taking the product of dark photon mass times kinetic mixing to be as large as possible given the constraints from solar dark photon emission shown in Fig.~\ref{fig:sun}.
As long as $m\lesssim10$~eV, the absorption rate of dark photons scales as $\epsilon^2 m^2$, so it is straightforward to rescale the sensitivity for arbitrary values of $\epsilon$ and $m$.

The S2-only analysis~\cite{XENON:2019gfn} by the XENON1T collaboration leads to the strongest constraints for dark radiation corresponding to a black body with $T\gtrsim10\,T_\odot$.
However, the lower threshold and background rate of the SENSEI experiment~\cite{SENSEI:2023zdf,SENSEI:2024yyt} gives it an edge for cooler dark galactic emission.
At their respective peak sensitivities, both experiments would be sensitive to a dark photon source of $\sim 10^{16}\,L_\odot$ at the center of our galaxy.
For reference, the Milky Way bulge has a total luminosity of $\sim 10^{10}\,L_\odot$ in visible photons.
Alternatively, a solar-luminosity dark star would need to be located at most $\sim 10^{-4}$~pc away for its dark starlight to be detectable.

In Fig.~\ref{fig:dark_galaxy} we also display the projected reach of full-scale SENSEI~\cite{SENSEI:2020dpa} (green dashes) and a futuristic 30~kg$\cdot$year silicon detector~\cite{2022arXiv220210518A} (green dots).
The lines represent the smallest accessible dark galaxy luminosities, corresponding to values of the kinetic mixing and dark photon mass near the current exclusion limits from solar emission shown in Fig.~\ref{fig:sun}.
Thus, the detection of a low-luminosity dark galaxy would most likely follow the detection of solar dark photon emission.
Since DM experiments are not sensitive to the direction of the dark photon flux, solar emission is effectively a foreground for the dark starlight signal.
Therefore, in the event of a detection with sufficient statistics, the DM-sourced events can at best be constrained by $N_{\rm DM}\leq 1.3\sqrt{N_{\odot}}$ at 90\% C.L.~in the presence of the solar foreground, as long as the latter is accurately modeled. 
We use this figure for our projections in Fig.~\ref{fig:dark_galaxy}.
The scaling of the dark starlight reach with exposure is more favorable than the solar dark photon one, since for the former the signal rate scales with $\epsilon^2$ as opposed to $\epsilon^4$ for the latter.

\subsection{Resonant surface conversion}

Detecting dark photons through volumetric absorption in dark matter detections has one major drawback: the lack of directional information.
Directionality is an effective tool reject experimental backgrounds and characterize any found signal. 
This is of obvious utility in the search for solar dark photon emission, where directionality would confirm the sun as the source of the signal. 
In the search for DM-sourced dark radiation, directionality allows for partial rejection the solar dark photon foreground, 
and ultimately to map the galactic emission if a signal is detected.
Though volumetric absorption cannot lead to directionality in an isotropic material, the new surface conversion process occurs only near the surface of the target material and is thus inherently directional.
As we now argue, this process can be exploited to develop a new class of experiments with spatial sensitivity to the dark radiation signal.

Resonant surface conversion occurs in materials that can support propagating longitudinal plasmons.
Examples are plasmas, (super and semi)conductors, noble gases for frequencies above the ionization energy\dots\ 
In this paper, we focus on conducting metals, where the photon longitudinal mode corresponds to collective motion of the conduction electrons within the material.
As a benchmark, we consider copper, which has a plasma frequency $\omega_p\simeq 10$~eV and whose high DC conductivity is useful experimentally, as we argue below.

We model the frequency-dependent (complex) conductivity of copper with a simple Drude-Sommerfeld free electron model
\begin{equation}
    \sigma(\omega) = \frac{\sigma_0}{1-i\omega\tau},
\end{equation}
where $\sigma_0$ is the DC conductivity and $\tau$ is the characteristic timescale between electron collisions that interrupt their free motion.
This translates into a frequency-dependent polarization tensor $\pi_T = -i\omega\sigma(\omega)$.
The plasma frequency is
\begin{equation}
    \omega_p = \sqrt{\frac{\sigma_0}{\tau}},
\end{equation}
and we also define the quality factor of the conductor as
\begin{equation}
    Q = \frac{\omega_p}{\mathrm{Re}\,\sigma(\omega_p)} = \frac{\sigma_0}{\omega_p} = \omega_p \tau = \sqrt{\sigma_0\tau}.
\end{equation}
A larger value of $Q$ corresponds to a larger DC conductivity but to a \textit{smaller} conductivity at the plasma frequency, which enhances the absorption rate in Eq.~\eqref{eq:absorption_rate_surface_resonance}.
Semiconductors like silicon have a much smaller quality factor at their plasma frequency, making their resonance much more challenging to exploit experimentally.

A Drude-Sommerfeld conductor is fully parametrized by $\omega_p$ and $Q$.
For copper, $\omega_p\simeq 10$~eV is largely independently of temperature and sample purity.
However, $Q$ is highly sensitive to impurities in the material and to temperature. 
The scattering time $\tau$ gets two main additive contributions, from electron scatterings with phonons and with impurities.
At room temperature, the phonon contribution dominates leading to a conductivity $\sigma_0(T=300\,\mathrm{K}) \simeq 6\times10^7\,\mathrm{S}/\mathrm{m}$, and thus a quality factor of $Q\simeq 400$.

At low temperatures, the phonon contribution to $\tau$ is well described by the Bloch-Gr\"uneisen formula~\cite{Mizutani_2001} which gives a contribution $\propto T^5$ to the resistivity.
The impurities contribution is largely independent of temperature and thus dominates at the very low temperatures of interest here.
In this case, the resistivity is usually characterized by the residual resistivity ratio (RRR)
\begin{equation}
    \mathrm{RRR} \equiv \frac{\rho_0(T=300\,\mathrm{K})}{\rho_0 (T=0\,\mathrm{K}) },
\end{equation}
where $\rho_0$ denotes the DC resistivity.
Ultra-pure copper has RRR values in the $100-400$ range.
This translates into quality factors in the $Q\simeq 10^4 - 10^5$ range at cryogenic temperatures.
Even if a larger value of $Q$ leads to a narrower frequency response, the frequency-integrated event rate stays constant, as we will see.
This is in contrast to searches of low-brightness astronomical sources, for which the limitation is to capture as large a fraction of the photons arriving at the telescope as possible.
In our case, there is a large amount of dark photons arriving at the detector even in a small frequency band, and the challenge is to convert as large a fraction of them as possible while keeping experimental backgrounds under control.
As we now argue, a larger quality factor facilitates this task by enabling the construction of larger targets.

To exploit the directionality of the signal, the target material can be arranged in thin plates of surface area $A$ and thickness $l$, with $l\ll \sqrt{A}$.
When the dark radiation is incident normally on the copper plate, the total event rate throughout its thickness due to resonant absorption can be found by integrating the frequency-dependent absorption rate in Eq.~\eqref{eq:absorption_rate_surface_resonance},
\begin{align}
    R_{\rm surf} &= 2 \int_{z=0}^{z=l}\mathop{\diff z}\int\mathop{\diff \omega} \frac{1}{\beta} \Gamma_{\rm surf} \frac{\diff\Phi}{\diff\omega} A \nonumber \\
    &\simeq 2 \epsilon^2m^2 \frac{\diff\Phi}{\diff\omega}\bigg\rvert_{\omega_p} \frac{Q}{\omega_p} A (1-\me^{-\omega_p l / Q}),
\end{align}
where we have assumed that $Q\gg 1$ and $m\ll\omega_p$ so that $\beta\simeq 1$.
We have used a narrow width approximation to perform the $\omega$ integral since the resonant conversion is highly concentrated around $\omega_p$.
For thin, high-purity copper slabs satisfying $\omega_p l / Q\ll 1$, the rate simplifies to 
\begin{equation}
    R_{\rm surf}\simeq 2 \epsilon^2 m^2 \frac{\diff\Phi}{\diff\omega}\bigg\rvert_{\omega_p} l A,
\end{equation}
independently of the quality factor.
Thus, the target volume $V=lA$ is the figure of merit to be maximized within the experimental constraints.

\begin{figure*}[t]
\centering
\centerline{\includegraphics[width=0.65\linewidth]{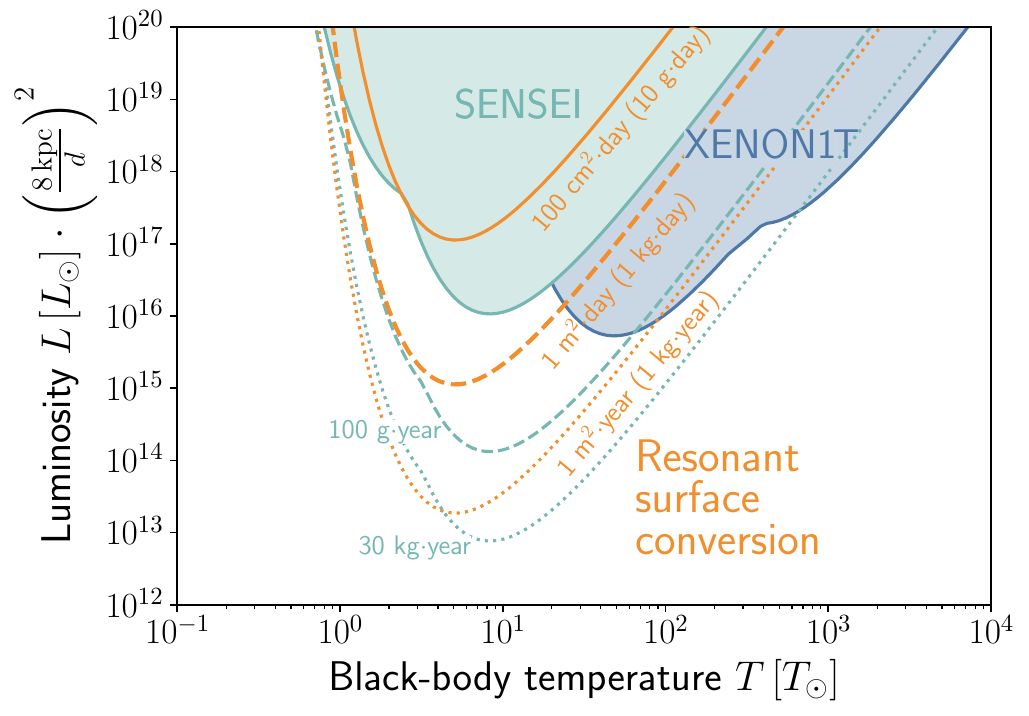}}
\caption{
Existing constraints and ability of proposed searches to detect the longitudinal dark photon emission from a dissipative galactic dark matter component. 
All limits are shown at 90\% C.L, and apply as long as $m\lesssim 10$~eV and $\epsilon$ is above the threshold required to yield at least $2.3$ signal events at each exposure.
Under those assumptions, the limits do not depend on the kinetic mixing parameter or dark photon mass due to the $\epsilon m$ and $\epsilon^2 m^2$ scaling of the galactic signal and solar background,  respectively (see text for details).
} 
\label{fig:dark_galaxy}
\end{figure*}

The strongest constraint on a target volume element is the feasibility of the readout of the energy deposited by the dark photon absorption.
In copper, this can be done bolometrically, similarly to how it is done in semiconductors for DM direct detection experiments like SuperCMDS~\cite{SuperCDMS:2016wui}.
The challenge is the higher specific heat capacity of copper compared to e.g.~silicon, driven by the electronic contribution that is absent in semiconductors at low temperatures.
The electronic contribution to the heat capacity scales linearly with temperature, in contrast to the phonon one that goes as $T^3$. 
This results in a huge difference between the specific heat capacity of copper and silicon at cryogenic temperatures,
\begin{align}
    C_{\rm Cu}(T=10\,\mathrm{mK}) \simeq 6\times 10^{11}\,\mathrm{eV}/\mathrm{g}\cdot\mathrm{K}, \nonumber\\
    C_{\rm Si}(T=10\,\mathrm{mK}) \simeq 1.6\times 10^{6}\,\mathrm{eV}/\mathrm{g}\cdot\mathrm{K}.
\end{align}
For copper, this limits the target volume given a minimum measurable temperature increase,
\begin{equation}
    lA \simeq \mathrm{mm}^3 \left( \frac{\mathrm{nK}}{\Delta T} \right) \left( \frac{\Delta E}{10\,\mathrm{eV}} \right) \left( \frac{10\,\mathrm{mK}}{T} \right).
\end{equation}
For nanokelvin calorimetry as can be achieved with cryogenically-operated transition-edge sensors~\cite{Chen:2021tap}, individual copper plates are limited to mm$^3$ volumes, or roughly $0.01$~g masses, operating at $10$~mK.
For this volume of copper, the calorimetric requirements are similar to what has been demonstrated in particle physics applications~\cite{Ricochet:2023yek} using larger semiconductor targets.
For our benchmark sensitivity projections, we assume a target thickness of $l=0.1$~mm and an area of $A=10\,\mathrm{mm}^2$.

We are now in a position to obtain a sensitivity estimate for a concrete dark radiation detector exploiting the surface conversion process in copper.
The target material consists of $N$ thin plates of $l=0.1$~mm thickness and surface area $A=10\,\mathrm{mm}^2$, for a total area of $A_{\rm tot}=NA$. 
These plates can be stacked into a compact experimental form factor.
We express the total experimental exposure in units of total surface area$\cdot$time, for a fixed plate thickness of $0.1$~mm, choosing benchmarks of $100\,\mathrm{cm}^2\cdot\mathrm{day}$, $1\,\mathrm{m}^2\cdot\mathrm{day}$, and $1\,\mathrm{m}^2\cdot\mathrm{year}$.
For easier comparison with volumetric absorption experiments, these correspond to $10\,\mathrm{g}\cdot\mathrm{day}$, $1\,\mathrm{kg}\cdot\mathrm{day}$, and $1\,\mathrm{kg}\cdot\mathrm{year}$ mass exposures, respectively.

The quality factor of copper is assumed to be $10^4 \lesssim Q \lesssim 10^5$, so that the thin dimension of the slab is smaller than the photon mean free path but the long dimensions are much larger than it.
Thus, the contribution to the resonant conversion on the thin edges of the plate is negligible compared to that on the plate face.
This generates the directional response of the detector: the resonant surface conversion rate is maximal when the dark radiation is incident normally to the plate face, and decreases as $\cos\theta$ for oblique incidence, where $\theta$ is the angle with the normal to the surface.

The reach of this setup is shown in Fig.~\ref{fig:sun} for solar dark photons.
We assume the search to be background free and calculate the $90\%$ C.L.~on the signal rate as $2.3$ divided by the exposure.
Fig.~\ref{fig:sun} shows that a just over m$^2\cdot$day experiment would start to test new parameter space for solar dark photons with masses $m\lesssim 10$~eV.
Larger exposures lead to comparable nominal sensitivities to silicon-based dark matter detectors.
This is as expected from Fig.~\ref{fig:GammaAS}, which shows that volume and surface absorption rates in Cu and Si are similar as long as the expected flux has no strong features in the $10-20$~eV range, as the solar one.
The key to delivering the forecasted sensitivity will be keeping the experiment background free, which will be aided by the directional sensitivity. 
Furthermore, since the signal source is localized in space, directionality will greatly enhance the robustness of the solar dark photon search, 
allowing for calibration with off-target runs, and unequivocal association of the signal with the sun in the event of a detection. 
This gives the surface-conversion copper experiment an edge compared to dark matter direct detection experiments, which are inherently non directional for the dark photon signal and thus more prone to signal contamination by backgrounds.

The sensitivity of the resonant surface conversion experiment to the dark starlight signal is shown in Fig.~\ref{fig:dark_galaxy}.
For the 100~cm$^2$ and 1m$^2\cdot$day exposures, no solar dark photon foreground is expected based on Fig.~\ref{fig:sun}, so we treat the dark galaxy search as background free. 
For a m$^2\cdot$year experiment, there could be a simultaneous detection of the solar and dark starlight emissions for the largest, currently unconstrained values of the kinetic mixing parameter. 
Thus, the projected reach to DM sources is calculated as $N_{\rm DM}\leq 1.3\sqrt{N_{\odot}\cdot\cos\alpha\cdot2/\pi}$ in the presence of the solar foreground, where $\alpha\simeq 60.2^\circ$ is the offset between the solar system and the galactic planes.
The geometric factors account for the suppression of the solar flux on the surface detector, assuming that it is permanently oriented towards the galactic center.

It is important to highlight that the directional nature of the resonant conversion strategy would allow for a data-driven subtraction of the solar foreground, since this component would be modulated on a yearly timescale if the detector is always pointed towards the galactic center.
Again the surface-conversion detection scheme is more robust than volumetric detection, this time in particular against possible mis-modeling of the solar dark photon flux.
Furthermore, if a sufficiently strong signal is found, the directionality would allow the emission profile of the dark matter in the galaxy to be mapped.

A resonant surface conversion experiment will test new regions of parameter space even with modest exposures, especially at lower temperatures corresponding to less energetic depositions for which low-threshold DM detectors suffer from unidentified backgrounds~\cite{Fuss:2022fxe}. 
A m$^2\cdot$year exposure  would result in sensitivity to $\sim10^{13}$ solar luminosities from the galactic center for emission profiles of a few $T_\odot$ (or equivalently, a single solar-luminosity compact source located $3$~mpc away).
For comparison, the luminosity of the Milky Way bulge, which is dominated by stars with $T\lesssim  T_\odot$, is $\sim10^{10}L_\odot$, and the luminosity rapidly increases with increased stellar temperature (and mass), making $10^{10}-10^{14}\,L_\odot$ a motivated target if the dark galaxy were to contain a similar number of stars to the visible one, just slightly hotter.

A priori, the directional dependence of the signal for a collimated source is simply $R_{\rm surf}\propto\cos\theta$, where $\theta$ is the angle between the direction of travel of the incoming dark radiation and the normal to the detector plate.
Of course, the detector array need not be equiplanar, and the different $10$~mm$^2$ modular target elements can be arranged to obtain different effective angular responses, and be dynamically rearranged to e.g.~track moving sources in the sky.
We leave further investigation of the full possibilities of this setup for the future.

\section{Summary and Conclusions}\label{sec:conclusions}

In this work, we have for the first time examined the detection prospects of a dark radiation signal originating from a dissipative dark matter sector.
In addition to showcasing the potential of dark matter experiments like XENON and SENSEI, we have identified a novel process based on resonant dark-photon-to-photon conversion at a conductor surface that offers a unique directional sensitivity to the signal.

A massive dark photon is assumed to mediate the dissipative interactions between dark matter particles and thus be the dark radiation quantum.
We focus on the detection of longitudinal modes, whose interaction rate within a visible medium is larger than that of transverse modes by a factor of $(\omega/m)^2$, where $\omega$ is the frequency of the incoming dark radiation and $m$ is the dark photon mass. We also argue that high-density dark matter processes favor longitudinal dark photon emission compared to transverse one.

We employ a two-state flavor formalism to describe the mixing between dark photons and longitudinal plasmons within a visible-sector conducting medium.
We identify two propagation eigenstates, a mostly dark and a mostly visible one.
For the mostly dark one, we recover the volumetric $\epsilon^2$-suppressed absorption rate found in Ref.~\cite{An:2013yfc}.
We also show that the visible, mostly plasmon one, which had been neglected in previous studies, is excited at the surface by the incoming dark photon wave and is absorbed at the same rate as transverse visible photons.
This leads to a strongly depth-dependent absorption rate when a dark photon enters a conducting medium from vacuum.
This effect is strongly enhanced when the frequency (not the mass) of the incoming dark radiation matches the plasma frequency of the visible medium, leading to a resonant absorption rate whose magnitude within a photon mean free path of the interface is similar to the volumetric process.

To exploit the surface resonant absorption process, we envision an experiment consisting of a series of thin plates made up of a good conductor.
The thickness of the plates is smaller than the mean free path of photons of frequency matching the plasma frequency of the material, while the other dimensions are larger.
This generates a $\cos \theta$ directional response of the detector.

We propose a concrete setup with copper as the target material.
Ultra-pure copper at cryogenic temperatures and with a residual resistivity ratio of $\sim$100 has a millimeter photon mean free path at its plasma frequency $\omega_p=10$~eV.
This constrains the thickness of the plates to be $\ll$~mm, and their area to be $\gg\mathrm{mm}^2$.
Due to the high electron-driven heat capacity of conductors, the volume of individual target modules is limited if the readout of the energy deposition is done bolometrically.
For copper cooled to $10$~mK, an energy deposition of $10$~eV in a sample of mm$^3$ volume generates a nK temperature increase, comparable to what can be measured with state-of-the-art transition-edge sensors.
An experimental array made of individual plates of $0.1\,\mathrm{mm}$ thickness and $10\,\mathrm{mm}^2$ surface area satisfies all the above requirements.

Many different models predict the existence of dissipative dynamics in the dark sector, with cosmological and astrophysical implications ranging from dark acoustic oscillations to the formation of compact objects, including dark stars powered by dark nuclear fusion, simple cooling on a long Kelvin-Helmholtz timescale, or other processes internal to the dark sector.
To remain agnostic about the microphysical origin of the signal, we parametrize its spectrum as a black body, and vary its temperature and total luminosity.
Motivated by a toy calculation of the massive dark photon luminosity of a sun-like dark star, which is dominantly longitudinal, we assume the dark radiation flux to be longitudinally polarized.
For other scenarios, our results apply to the longitudinal component of the dark radiation flux.

Aside from experimental backgrounds, the main foreground in the search for a dark-matter-sourced dark radiation signal is the production of dark photons in the sun through photon-to-dark-photon conversion in the solar plasma.
Of course, detecting solar dark photon emission is a goal on its own in the search for new physics, and further motivates designing experiments sensitive to longitudinal dark photons independently of their potential role as mediators of dark matter self interactions.
Indeed, compared to the absorption in DM experiments, the directional detection method that we propose here would help confirm that a hypothetical signal does, in fact, originate from the sun or the dark galaxy.

Our main results are summarized in Figs.~\ref{fig:sun} and~\ref{fig:dark_galaxy}.
They show the sensitivity to the solar and DM-sourced dark radiation signal from the SENSEI and XENON1T experiments, along with projections for future silicon-based detectors and for an experiment exploiting the novel resonant surface conversion process.
The unique directional sensitivity of our proposal will be a useful tool to reject experimental backgrounds as well as to separate the solar foreground should a signal be found.
This translates into an excellent discovery potential to dark galactic luminosities down to $10^{13}\,L_\odot$ with significantly smaller exposures than would be needed in traditional dark matter experiments.

In a way, our proposal is reminiscent of the helioscopes~\cite{Redondo:2008aa,Schwarz:2015lqa} that have been used to look for axion and solar dark photon emission (see~\cite{OShea:2023gqn} for a recent discussion of the prospects of the IAXO experiment).
The sensitivity of these experiments is limited by the fact that they are only sensitive to transverse dark photons, whose interaction rate is suppressed and is resonantly enhanced only when the detection medium plasma mass matches the dark photon mass, thus requiring time-consuming scanning over this a priori unknown model parameter.
In contrast, our novel resonant conversion process is sensitive to longitudinal modes and does not require scanning over unknown model parameters.
These two advantages makes the surface conversion scheme a much more promising setup to detect massive dark photon radiation. 

Our findings motivate further studies to explore the potential of existing experiments and new proposals to detect the dark radiation signal from dissipative dark matter.
A more realistic characterization of the expected signal arising in concrete models, including the emission from dark stars and other dark matter compact objects, is necessary to inform experimental designs.
For our copper-based resonant surface conversion proposal, fleshing out a readout strategy as well as characterizing possible backgrounds are imperative, as is designing a geometrical layout of the plates to maximize the potential of the directional sensitivity.
Furthermore, it is worth exploring other materials beyond copper that can exploit the resonant conversion process, including semiconductors and superconductors. 
Finally, the non-resonant component of the surface conversion warrants further examination, since it could offer access to lower dark photon frequencies than the detector material's plasma frequency.

We hope that this work initiates the experimental \textit{dark astronomy} program, a uniquely powerful search for radiative signals from dissipative dark sectors, to probe the nature of dark matter.

\section*{Acknowledgements}
We are grateful to Miriam Diamond, Ziqing Hong, Joerg Jaeckel, Yoni Kahn, Tongyang Lin, Chris Matzner, Georg Raffelt, Javier Redondo, Katelin Schutz, and John Sipe for useful discussions pertaining to this work.
The work of GA and DC was supported in part by Discovery Grants from the Natural Sciences and Engineering Research Council of Canada, the Canada Research Chair program, the Alfred P. Sloan Foundation, the Ontario Early Researcher Award, and the University of Toronto McLean Award.
GA thanks the Munich Institute for Astro-, Particle and BioPhysics (MIAPbP) funded by the DFG under (project ID EXC-2094-390783311), and the Mainz Institute for Theoretical Physics (MITP) of the Cluster of Excellence PRISMA$^+$ (project ID 39083149) for their hospitality while completing parts of this work.
This research was enabled in part by support provided by the Digital Research Alliance of Canada.
Numerical software used includes \texttt{Numpy}~\cite{harris2020array}, \texttt{Scipy}~\cite{2020SciPy-NMeth}, and \texttt{Matplotlib}~\cite{Hunter:2007}.

\bibliography{bibliography}

\providecommand{\href}[2]{#2}\begingroup\raggedright\begin{thebibliography}{100}

\bibitem{Akerib:2022ort}
D.~S. Akerib {\em et~al.}, ``{Snowmass2021 Cosmic Frontier Dark Matter Direct Detection to the Neutrino Fog},'' in {\em {Snowmass 2021}}.
\newblock 3, 2022.
\newblock \href{https://arxiv.org/abs/2203.08084}{{\ttfamily arXiv:2203.08084 [hep-ex]}}.

\bibitem{Essig:2022dfa}
R.~Essig {\em et~al.}, ``{Snowmass2021 Cosmic Frontier: The landscape of low-threshold dark matter direct detection in the next decade},'' in {\em {Snowmass 2021}}.
\newblock 3, 2022.
\newblock \href{https://arxiv.org/abs/2203.08297}{{\ttfamily arXiv:2203.08297 [hep-ph]}}.

\bibitem{Knapen:2017xzo}
S.~Knapen, T.~Lin, and K.~M. Zurek, ``{Light Dark Matter: Models and Constraints},'' \href{https://dx.doi.org/10.1103/PhysRevD.96.115021}{{\em Phys. Rev. D} {\bfseries 96} no.~11, (2017) 115021}, \href{https://arxiv.org/abs/1709.07882}{{\ttfamily arXiv:1709.07882 [hep-ph]}}.

\bibitem{Antel:2023hkf}
C.~Antel {\em et~al.}, ``{Feebly-interacting particles: FIPs 2022 Workshop Report},'' \href{https://dx.doi.org/10.1140/epjc/s10052-023-12168-5}{{\em Eur. Phys. J. C} {\bfseries 83} no.~12, (2023) 1122}, \href{https://arxiv.org/abs/2305.01715}{{\ttfamily arXiv:2305.01715 [hep-ph]}}.

\bibitem{Leane:2020liq}
R.~K. Leane, ``{Indirect Detection of Dark Matter in the Galaxy},'' in {\em {3rd World Summit on Exploring the Dark Side of the Universe}}, pp.~203--228.
\newblock 2020.
\newblock \href{https://arxiv.org/abs/2006.00513}{{\ttfamily arXiv:2006.00513 [hep-ph]}}.

\bibitem{Bertone:2019irm}
G.~Bertone {\em et~al.}, ``{Gravitational wave probes of dark matter: challenges and opportunities},'' \href{https://dx.doi.org/10.21468/SciPostPhysCore.3.2.007}{{\em SciPost Phys. Core} {\bfseries 3} (2020) 007}, \href{https://arxiv.org/abs/1907.10610}{{\ttfamily arXiv:1907.10610 [astro-ph.CO]}}.

\bibitem{Arguelles:2019ouk}
C.~A. Arg\"uelles, A.~Diaz, A.~Kheirandish, A.~Olivares-Del-Campo, I.~Safa, and A.~C. Vincent, ``{Dark matter annihilation to neutrinos},'' \href{https://dx.doi.org/10.1103/RevModPhys.93.035007}{{\em Rev. Mod. Phys.} {\bfseries 93} no.~3, (2021) 035007}, \href{https://arxiv.org/abs/1912.09486}{{\ttfamily arXiv:1912.09486 [hep-ph]}}.

\bibitem{PAMELA:2008gwm}
{\bfseries PAMELA} Collaboration, O.~Adriani {\em et~al.}, ``{An anomalous positron abundance in cosmic rays with energies 1.5-100 GeV},'' \href{https://dx.doi.org/10.1038/nature07942}{{\em Nature} {\bfseries 458} (2009) 607--609}, \href{https://arxiv.org/abs/0810.4995}{{\ttfamily arXiv:0810.4995 [astro-ph]}}.

\bibitem{Hooper:2010mq}
D.~Hooper and L.~Goodenough, ``{Dark Matter Annihilation in The Galactic Center As Seen by the Fermi Gamma Ray Space Telescope},'' \href{https://dx.doi.org/10.1016/j.physletb.2011.02.029}{{\em Phys. Lett. B} {\bfseries 697} (2011) 412--428}, \href{https://arxiv.org/abs/1010.2752}{{\ttfamily arXiv:1010.2752 [hep-ph]}}.

\bibitem{Siegert:2015knp}
T.~Siegert, R.~Diehl, G.~Khachatryan, M.~G.~H. Krause, F.~Guglielmetti, J.~Greiner, A.~W. Strong, and X.~Zhang, ``{Gamma-ray spectroscopy of Positron Annihilation in the Milky Way},'' \href{https://dx.doi.org/10.1051/0004-6361/201527510}{{\em Astron. Astrophys.} {\bfseries 586} (2016) A84}, \href{https://arxiv.org/abs/1512.00325}{{\ttfamily arXiv:1512.00325 [astro-ph.HE]}}.

\bibitem{Cui:2016ppb}
M.-Y. Cui, Q.~Yuan, Y.-L.~S. Tsai, and Y.-Z. Fan, ``{Possible dark matter annihilation signal in the AMS-02 antiproton data},'' \href{https://dx.doi.org/10.1103/PhysRevLett.118.191101}{{\em Phys. Rev. Lett.} {\bfseries 118} no.~19, (2017) 191101}, \href{https://arxiv.org/abs/1610.03840}{{\ttfamily arXiv:1610.03840 [astro-ph.HE]}}.

\bibitem{Cuoco:2016eej}
A.~Cuoco, M.~Kr\"amer, and M.~Korsmeier, ``{Novel Dark Matter Constraints from Antiprotons in Light of AMS-02},'' \href{https://dx.doi.org/10.1103/PhysRevLett.118.191102}{{\em Phys. Rev. Lett.} {\bfseries 118} no.~19, (2017) 191102}, \href{https://arxiv.org/abs/1610.03071}{{\ttfamily arXiv:1610.03071 [astro-ph.HE]}}.

\bibitem{AMS_AntiHe}
V.~A. Choutko. AMS Days at La Palma, Spain (2018).

\bibitem{Cyr-Racine:2015ihg}
F.-Y. Cyr-Racine, K.~Sigurdson, J.~Zavala, T.~Bringmann, M.~Vogelsberger, and C.~Pfrommer, ``{ETHOS\textemdash{}an effective theory of structure formation: From dark particle physics to the matter distribution of the Universe},'' \href{https://dx.doi.org/10.1103/PhysRevD.93.123527}{{\em Phys. Rev. D} {\bfseries 93} no.~12, (2016) 123527}, \href{https://arxiv.org/abs/1512.05344}{{\ttfamily arXiv:1512.05344 [astro-ph.CO]}}.

\bibitem{Tulin:2017ara}
S.~Tulin and H.-B. Yu, ``{Dark Matter Self-interactions and Small Scale Structure},'' \href{https://dx.doi.org/10.1016/j.physrep.2017.11.004}{{\em Phys. Rept.} {\bfseries 730} (2018) 1--57}, \href{https://arxiv.org/abs/1705.02358}{{\ttfamily arXiv:1705.02358 [hep-ph]}}.

\bibitem{Bernal:2015ova}
N.~Bernal, X.~Chu, C.~Garcia-Cely, T.~Hambye, and B.~Zaldivar, ``{Production Regimes for Self-Interacting Dark Matter},'' \href{https://dx.doi.org/10.1088/1475-7516/2016/03/018}{{\em JCAP} {\bfseries 03} (2016) 018}, \href{https://arxiv.org/abs/1510.08063}{{\ttfamily arXiv:1510.08063 [hep-ph]}}.

\bibitem{Ackerman:2008kmp}
L.~Ackerman, M.~R. Buckley, S.~M. Carroll, and M.~Kamionkowski, ``{Dark Matter and Dark Radiation},'' \href{https://dx.doi.org/10.1103/PhysRevD.79.023519}{{\em Phys. Rev. D} {\bfseries 79} (2009) 023519}, \href{https://arxiv.org/abs/0810.5126}{{\ttfamily arXiv:0810.5126 [hep-ph]}}.

\bibitem{Randall:2008ppe}
S.~W. Randall, M.~Markevitch, D.~Clowe, A.~H. Gonzalez, and M.~Bradac, ``{Constraints on the Self-Interaction Cross-Section of Dark Matter from Numerical Simulations of the Merging Galaxy Cluster 1E 0657-56},'' \href{https://dx.doi.org/10.1086/587859}{{\em Astrophys. J.} {\bfseries 679} (2008) 1173--1180}, \href{https://arxiv.org/abs/0704.0261}{{\ttfamily arXiv:0704.0261 [astro-ph]}}.

\bibitem{Tucker-Smith:2001myb}
D.~Tucker-Smith and N.~Weiner, ``{Inelastic dark matter},'' \href{https://dx.doi.org/10.1103/PhysRevD.64.043502}{{\em Phys. Rev. D} {\bfseries 64} (2001) 043502}, \href{https://arxiv.org/abs/hep-ph/0101138}{{\ttfamily arXiv:hep-ph/0101138}}.

\bibitem{Arkani-Hamed:2008hhe}
N.~Arkani-Hamed, D.~P. Finkbeiner, T.~R. Slatyer, and N.~Weiner, ``{A Theory of Dark Matter},'' \href{https://dx.doi.org/10.1103/PhysRevD.79.015014}{{\em Phys. Rev. D} {\bfseries 79} (2009) 015014}, \href{https://arxiv.org/abs/0810.0713}{{\ttfamily arXiv:0810.0713 [hep-ph]}}.

\bibitem{Schutz:2014nka}
K.~Schutz and T.~R. Slatyer, ``{Self-Scattering for Dark Matter with an Excited State},'' \href{https://dx.doi.org/10.1088/1475-7516/2015/01/021}{{\em JCAP} {\bfseries 01} (2015) 021}, \href{https://arxiv.org/abs/1409.2867}{{\ttfamily arXiv:1409.2867 [hep-ph]}}.

\bibitem{Blennow:2016gde}
M.~Blennow, S.~Clementz, and J.~Herrero-Garcia, ``{Self-interacting inelastic dark matter: A viable solution to the small scale structure problems},'' \href{https://dx.doi.org/10.1088/1475-7516/2017/03/048}{{\em JCAP} {\bfseries 03} (2017) 048}, \href{https://arxiv.org/abs/1612.06681}{{\ttfamily arXiv:1612.06681 [hep-ph]}}.

\bibitem{Zhang:2016dck}
Y.~Zhang, ``{Self-interacting Dark Matter Without Direct Detection Constraints},'' \href{https://dx.doi.org/10.1016/j.dark.2016.12.003}{{\em Phys. Dark Univ.} {\bfseries 15} (2017) 82--89}, \href{https://arxiv.org/abs/1611.03492}{{\ttfamily arXiv:1611.03492 [hep-ph]}}.

\bibitem{2019MNRAS.484.5437V}
M.~{Vogelsberger}, J.~{Zavala}, K.~{Schutz}, and T.~R. {Slatyer}, ``{Evaporating the Milky Way halo and its satellites with inelastic self-interacting dark matter},'' \href{https://dx.doi.org/10.1093/mnras/stz340}{{\em \mnras} {\bfseries 484} no.~4, (Apr., 2019) 5437--5452}, \href{https://arxiv.org/abs/1805.03203}{{\ttfamily arXiv:1805.03203 [astro-ph.GA]}}.

\bibitem{Alvarez:2019nwt}
G.~Alvarez and H.-B. Yu, ``{Astrophysical probes of inelastic dark matter with a light mediator},'' \href{https://dx.doi.org/10.1103/PhysRevD.101.043002}{{\em Phys. Rev. D} {\bfseries 101} no.~4, (2020) 043002}, \href{https://arxiv.org/abs/1911.11114}{{\ttfamily arXiv:1911.11114 [hep-ph]}}.

\bibitem{2021MNRAS.500.1531C}
K.~T.~E. {Chua}, K.~{Dibert}, M.~{Vogelsberger}, and J.~{Zavala}, ``{The impact of inelastic self-interacting dark matter on the dark matter structure of a Milky Way halo},'' \href{https://dx.doi.org/10.1093/mnras/staa3315}{{\em \mnras} {\bfseries 500} no.~1, (Jan., 2021) 1531--1546}, \href{https://arxiv.org/abs/2010.08562}{{\ttfamily arXiv:2010.08562 [astro-ph.GA]}}.

\bibitem{2023MNRAS.524..288O}
S.~{O'Neil}, M.~{Vogelsberger}, {\em et~al.}, ``{Endothermic self-interacting dark matter in Milky Way-like dark matter haloes},'' \href{https://dx.doi.org/10.1093/mnras/stad1850}{{\em \mnras} {\bfseries 524} no.~1, (Sept., 2023) 288--306}, \href{https://arxiv.org/abs/2210.16328}{{\ttfamily arXiv:2210.16328 [astro-ph.GA]}}.

\bibitem{Leonard:2024mqo}
A.~Leonard, S.~O'Neil, X.~Shen, M.~Vogelsberger, O.~Rosenstein, H.~Shangguan, Y.~Teng, and J.~Hu, ``{Varying primordial state fractions in exo- and endothermic SIDM simulations of Milky Way-mass haloes},'' \href{https://dx.doi.org/10.1093/mnras/stae1270}{{\em Mon. Not. Roy. Astron. Soc.} {\bfseries 531} no.~1, (2024) 1440--1453}, \href{https://arxiv.org/abs/2401.13727}{{\ttfamily arXiv:2401.13727 [astro-ph.CO]}}.

\bibitem{Detmold:2014qqa}
W.~Detmold, M.~McCullough, and A.~Pochinsky, ``{Dark Nuclei I: Cosmology and Indirect Detection},'' \href{https://dx.doi.org/10.1103/PhysRevD.90.115013}{{\em Phys. Rev. D} {\bfseries 90} no.~11, (2014) 115013}, \href{https://arxiv.org/abs/1406.2276}{{\ttfamily arXiv:1406.2276 [hep-ph]}}.

\bibitem{Wise:2014jva}
M.~B. Wise and Y.~Zhang, ``{Stable Bound States of Asymmetric Dark Matter},'' \href{https://dx.doi.org/10.1103/PhysRevD.90.055030}{{\em Phys. Rev. D} {\bfseries 90} no.~5, (2014) 055030}, \href{https://arxiv.org/abs/1407.4121}{{\ttfamily arXiv:1407.4121 [hep-ph]}}. [Erratum: Phys.Rev.D 91, 039907 (2015)].

\bibitem{Krnjaic:2014xza}
G.~Krnjaic and K.~Sigurdson, ``{Big Bang Darkleosynthesis},'' \href{https://dx.doi.org/10.1016/j.physletb.2015.11.001}{{\em Phys. Lett. B} {\bfseries 751} (2015) 464--468}, \href{https://arxiv.org/abs/1406.1171}{{\ttfamily arXiv:1406.1171 [hep-ph]}}.

\bibitem{Gresham:2018anj}
M.~I. Gresham, H.~K. Lou, and K.~M. Zurek, ``{Astrophysical Signatures of Asymmetric Dark Matter Bound States},'' \href{https://dx.doi.org/10.1103/PhysRevD.98.096001}{{\em Phys. Rev. D} {\bfseries 98} no.~9, (2018) 096001}, \href{https://arxiv.org/abs/1805.04512}{{\ttfamily arXiv:1805.04512 [hep-ph]}}.

\bibitem{Alves:2009nf}
D.~S.~M. Alves, S.~R. Behbahani, P.~Schuster, and J.~G. Wacker, ``{Composite Inelastic Dark Matter},'' \href{https://dx.doi.org/10.1016/j.physletb.2010.08.006}{{\em Phys. Lett. B} {\bfseries 692} (2010) 323--326}, \href{https://arxiv.org/abs/0903.3945}{{\ttfamily arXiv:0903.3945 [hep-ph]}}.

\bibitem{Boddy:2014yra}
K.~K. Boddy, J.~L. Feng, M.~Kaplinghat, and T.~M.~P. Tait, ``{Self-Interacting Dark Matter from a Non-Abelian Hidden Sector},'' \href{https://dx.doi.org/10.1103/PhysRevD.89.115017}{{\em Phys. Rev. D} {\bfseries 89} no.~11, (2014) 115017}, \href{https://arxiv.org/abs/1402.3629}{{\ttfamily arXiv:1402.3629 [hep-ph]}}.

\bibitem{Cline:2013zca}
J.~M. Cline, Z.~Liu, G.~D. Moore, and W.~Xue, ``{Composite strongly interacting dark matter},'' \href{https://dx.doi.org/10.1103/PhysRevD.90.015023}{{\em Phys. Rev. D} {\bfseries 90} no.~1, (2014) 015023}, \href{https://arxiv.org/abs/1312.3325}{{\ttfamily arXiv:1312.3325 [hep-ph]}}.

\bibitem{Alonso-Alvarez:2023rjq}
G.~Alonso-\'Alvarez, R.~Cao, J.~M. Cline, K.~Moorthy, and T.~Xiao, ``{Nonabelian kinetic mixing in a confining phase: a framework for composite dark photons},'' \href{https://dx.doi.org/10.1007/JHEP02(2024)017}{{\em JHEP} {\bfseries 02} (2024) 017}, \href{https://arxiv.org/abs/2309.13105}{{\ttfamily arXiv:2309.13105 [hep-ph]}}.

\bibitem{Kaplan:2009de}
D.~E. Kaplan, G.~Z. Krnjaic, K.~R. Rehermann, and C.~M. Wells, ``{Atomic Dark Matter},'' \href{https://dx.doi.org/10.1088/1475-7516/2010/05/021}{{\em JCAP} {\bfseries 05} (2010) 021}, \href{https://arxiv.org/abs/0909.0753}{{\ttfamily arXiv:0909.0753 [hep-ph]}}.

\bibitem{Fan:2013yva}
J.~Fan, A.~Katz, L.~Randall, and M.~Reece, ``{Double-Disk Dark Matter},'' \href{https://dx.doi.org/10.1016/j.dark.2013.07.001}{{\em Phys. Dark Univ.} {\bfseries 2} (2013) 139--156}, \href{https://arxiv.org/abs/1303.1521}{{\ttfamily arXiv:1303.1521 [astro-ph.CO]}}.

\bibitem{Foot:2014uba}
R.~Foot and S.~Vagnozzi, ``{Dissipative hidden sector dark matter},'' \href{https://dx.doi.org/10.1103/PhysRevD.91.023512}{{\em Phys. Rev. D} {\bfseries 91} (2015) 023512}, \href{https://arxiv.org/abs/1409.7174}{{\ttfamily arXiv:1409.7174 [hep-ph]}}.

\bibitem{Chang:2018bgx}
J.~H. Chang, D.~Egana-Ugrinovic, R.~Essig, and C.~Kouvaris, ``{Structure Formation and Exotic Compact Objects in a Dissipative Dark Sector},'' \href{https://dx.doi.org/10.1088/1475-7516/2019/03/036}{{\em JCAP} {\bfseries 03} (2019) 036}, \href{https://arxiv.org/abs/1812.07000}{{\ttfamily arXiv:1812.07000 [hep-ph]}}.

\bibitem{Chacko:2005pe}
Z.~Chacko, H.-S. Goh, and R.~Harnik, ``{The Twin Higgs: Natural electroweak breaking from mirror symmetry},'' \href{https://dx.doi.org/10.1103/PhysRevLett.96.231802}{{\em Phys. Rev. Lett.} {\bfseries 96} (2006) 231802}, \href{https://arxiv.org/abs/hep-ph/0506256}{{\ttfamily arXiv:hep-ph/0506256}}.

\bibitem{Kolb:1985bf}
E.~W. Kolb, D.~Seckel, and M.~S. Turner, ``{The Shadow World},'' \href{https://dx.doi.org/10.1038/314415a0}{{\em Nature} {\bfseries 314} (1985) 415--419}.

\bibitem{Glashow:1985ud}
S.~L. Glashow, ``{Positronium Versus the Mirror Universe},'' \href{https://dx.doi.org/10.1016/0370-2693(86)90540-X}{{\em Phys. Lett. B} {\bfseries 167} (1986) 35--36}.

\bibitem{Essig:2018pzq}
R.~Essig, S.~D. Mcdermott, H.-B. Yu, and Y.-M. Zhong, ``{Constraining Dissipative Dark Matter Self-Interactions},'' \href{https://dx.doi.org/10.1103/PhysRevLett.123.121102}{{\em Phys. Rev. Lett.} {\bfseries 123} no.~12, (2019) 121102}, \href{https://arxiv.org/abs/1809.01144}{{\ttfamily arXiv:1809.01144 [hep-ph]}}.

\bibitem{Huo:2019yhk}
R.~Huo, H.-B. Yu, and Y.-M. Zhong, ``{The Structure of Dissipative Dark Matter Halos},'' \href{https://dx.doi.org/10.1088/1475-7516/2020/06/051}{{\em JCAP} {\bfseries 06} (2020) 051}, \href{https://arxiv.org/abs/1912.06757}{{\ttfamily arXiv:1912.06757 [astro-ph.CO]}}.

\bibitem{Shen:2021frv}
X.~Shen, P.~F. Hopkins, L.~Necib, F.~Jiang, M.~Boylan-Kolchin, and A.~Wetzel, ``{Dissipative dark matter on FIRE \textendash{} I. Structural and kinematic properties of dwarf galaxies},'' \href{https://dx.doi.org/10.1093/mnras/stab2042}{{\em Mon. Not. Roy. Astron. Soc.} {\bfseries 506} no.~3, (2021) 4421--4445}, \href{https://arxiv.org/abs/2102.09580}{{\ttfamily arXiv:2102.09580 [astro-ph.GA]}}.

\bibitem{Shen:2022opd}
X.~Shen, P.~F. Hopkins, L.~Necib, F.~Jiang, M.~Boylan-Kolchin, and A.~Wetzel, ``{Dissipative Dark Matter on FIRE. II. Observational Signatures and Constraints from Local Dwarf Galaxies},'' \href{https://dx.doi.org/10.3847/1538-4357/ad2fb1}{{\em Astrophys. J.} {\bfseries 966} no.~1, (2024) 131}, \href{https://arxiv.org/abs/2206.05327}{{\ttfamily arXiv:2206.05327 [astro-ph.GA]}}.

\bibitem{Roy:2023zar}
S.~Roy, X.~Shen, M.~Lisanti, D.~Curtin, N.~Murray, and P.~F. Hopkins, ``{Simulating Atomic Dark Matter in Milky Way Analogs},'' \href{https://dx.doi.org/10.3847/2041-8213/ace2c8}{{\em Astrophys. J. Lett.} {\bfseries 954} no.~2, (2023) L40}, \href{https://arxiv.org/abs/2304.09878}{{\ttfamily arXiv:2304.09878 [astro-ph.GA]}}.

\bibitem{Gemmell:2023trd}
C.~Gemmell, S.~Roy, X.~Shen, D.~Curtin, M.~Lisanti, N.~Murray, and P.~F. Hopkins, ``{Dissipative Dark Substructure: The Consequences of Atomic Dark Matter on Milky Way Analog Subhalos},'' \href{https://dx.doi.org/10.3847/1538-4357/ad3823}{{\em Astrophys. J.} {\bfseries 967} no.~1, (2024) 21}, \href{https://arxiv.org/abs/2311.02148}{{\ttfamily arXiv:2311.02148 [astro-ph.GA]}}.

\bibitem{Roy:2024bcu}
S.~Roy, X.~Shen, J.~Barron, M.~Lisanti, D.~Curtin, N.~Murray, and P.~F. Hopkins, ``{Aggressively-Dissipative Dark Dwarfs: The Effects of Atomic Dark Matter on the Inner Densities of Isolated Dwarf Galaxies},'' \href{https://arxiv.org/abs/2408.15317}{{\ttfamily arXiv:2408.15317 [astro-ph.GA]}}.

\bibitem{2022ApJ...934..120R}
M.~{Ryan}, J.~{Gurian}, S.~{Shandera}, and D.~{Jeong}, ``{Molecular Chemistry for Dark Matter},'' \href{https://dx.doi.org/10.3847/1538-4357/ac75ef}{{\em \apj} {\bfseries 934} no.~2, (Aug., 2022) 120}, \href{https://arxiv.org/abs/2106.13245}{{\ttfamily arXiv:2106.13245 [astro-ph.CO]}}.

\bibitem{Buckley:2017ttd}
M.~R. Buckley and A.~DiFranzo, ``{Collapsed Dark Matter Structures},'' \href{https://dx.doi.org/10.1103/PhysRevLett.120.051102}{{\em Phys. Rev. Lett.} {\bfseries 120} no.~5, (2018) 051102}, \href{https://arxiv.org/abs/1707.03829}{{\ttfamily arXiv:1707.03829 [hep-ph]}}.

\bibitem{2022ApJ...939L..12G}
J.~{Gurian}, M.~{Ryan}, S.~{Schon}, D.~{Jeong}, and S.~{Shandera}, ``{A Lower Bound on the Mass of Compact Objects from Dissipative Dark Matter},'' \href{https://dx.doi.org/10.3847/2041-8213/ac997c}{{\em \apjl} {\bfseries 939} no.~1, (Nov., 2022) L12}, \href{https://arxiv.org/abs/2209.00064}{{\ttfamily arXiv:2209.00064 [astro-ph.CO]}}.

\bibitem{DAmico:2017lqj}
G.~D'Amico, P.~Panci, A.~Lupi, S.~Bovino, and J.~Silk, ``{Massive Black Holes from Dissipative Dark Matter},'' \href{https://dx.doi.org/10.1093/mnras/stx2419}{{\em Mon. Not. Roy. Astron. Soc.} {\bfseries 473} no.~1, (2018) 328--335}, \href{https://arxiv.org/abs/1707.03419}{{\ttfamily arXiv:1707.03419 [astro-ph.CO]}}.

\bibitem{Shandera:2018xkn}
S.~Shandera, D.~Jeong, and H.~S.~G. Gebhardt, ``{Gravitational Waves from Binary Mergers of Subsolar Mass Dark Black Holes},'' \href{https://dx.doi.org/10.1103/PhysRevLett.120.241102}{{\em Phys. Rev. Lett.} {\bfseries 120} no.~24, (2018) 241102}, \href{https://arxiv.org/abs/1802.08206}{{\ttfamily arXiv:1802.08206 [astro-ph.CO]}}.

\bibitem{Choquette:2018lvq}
J.~Choquette, J.~M. Cline, and J.~M. Cornell, ``{Early formation of supermassive black holes via dark matter self-interactions},'' \href{https://dx.doi.org/10.1088/1475-7516/2019/07/036}{{\em JCAP} {\bfseries 07} (2019) 036}, \href{https://arxiv.org/abs/1812.05088}{{\ttfamily arXiv:1812.05088 [astro-ph.CO]}}.

\bibitem{Latif:2018kqv}
M.~A. Latif, A.~Lupi, D.~R.~G. Schleicher, G.~D'Amico, P.~Panci, and S.~Bovino, ``{Black hole formation in the context of dissipative dark matter},'' \href{https://dx.doi.org/10.1093/mnras/stz608}{{\em Mon. Not. Roy. Astron. Soc.} {\bfseries 485} no.~3, (2019) 3352--3359}, \href{https://arxiv.org/abs/1812.03104}{{\ttfamily arXiv:1812.03104 [astro-ph.CO]}}.

\bibitem{Ryan:2022hku}
M.~Ryan and D.~Radice, ``{Exotic compact objects: The dark white dwarf},'' \href{https://dx.doi.org/10.1103/PhysRevD.105.115034}{{\em Phys. Rev. D} {\bfseries 105} no.~11, (2022) 115034}, \href{https://arxiv.org/abs/2201.05626}{{\ttfamily arXiv:2201.05626 [astro-ph.HE]}}.

\bibitem{Hippert:2021fch}
M.~Hippert, J.~Setford, H.~Tan, D.~Curtin, J.~Noronha-Hostler, and N.~Yunes, ``{Mirror neutron stars},'' \href{https://dx.doi.org/10.1103/PhysRevD.106.035025}{{\em Phys. Rev. D} {\bfseries 106} no.~3, (2022) 035025}, \href{https://arxiv.org/abs/2103.01965}{{\ttfamily arXiv:2103.01965 [astro-ph.HE]}}.

\bibitem{Curtin:2019ngc}
D.~Curtin and J.~Setford, ``{Signatures of Mirror Stars},'' \href{https://dx.doi.org/10.1007/JHEP03(2020)041}{{\em JHEP} {\bfseries 03} (2020) 041}, \href{https://arxiv.org/abs/1909.04072}{{\ttfamily arXiv:1909.04072 [hep-ph]}}.

\bibitem{Curtin:2019lhm}
D.~Curtin and J.~Setford, ``{How To Discover Mirror Stars},'' \href{https://dx.doi.org/10.1016/j.physletb.2020.135391}{{\em Phys. Lett. B} {\bfseries 804} (2020) 135391}, \href{https://arxiv.org/abs/1909.04071}{{\ttfamily arXiv:1909.04071 [hep-ph]}}.

\bibitem{Mohapatra:1996yy}
R.~N. Mohapatra and V.~L. Teplitz, ``{Structures in the mirror universe},'' \href{https://dx.doi.org/10.1086/303762}{{\em Astrophys. J.} {\bfseries 478} (1997) 29--38}, \href{https://arxiv.org/abs/astro-ph/9603049}{{\ttfamily arXiv:astro-ph/9603049}}.

\bibitem{Foot:1999hm}
R.~Foot, ``{Have mirror stars been observed?},'' \href{https://dx.doi.org/10.1016/S0370-2693(99)00230-0}{{\em Phys. Lett. B} {\bfseries 452} (1999) 83--86}, \href{https://arxiv.org/abs/astro-ph/9902065}{{\ttfamily arXiv:astro-ph/9902065}}.

\bibitem{Berezhiani:2003xm}
Z.~Berezhiani, ``{Mirror world and its cosmological consequences},'' \href{https://dx.doi.org/10.1142/S0217751X04020075}{{\em Int. J. Mod. Phys. A} {\bfseries 19} (2004) 3775--3806}, \href{https://arxiv.org/abs/hep-ph/0312335}{{\ttfamily arXiv:hep-ph/0312335}}.

\bibitem{Holdom:1985ag}
B.~Holdom, ``{Two U(1)'s and Epsilon Charge Shifts},'' \href{https://dx.doi.org/10.1016/0370-2693(86)91377-8}{{\em Phys. Lett. B} {\bfseries 166} (1986) 196--198}.

\bibitem{An:2013yua}
H.~An, M.~Pospelov, and J.~Pradler, ``{Dark Matter Detectors as Dark Photon Helioscopes},'' \href{https://dx.doi.org/10.1103/PhysRevLett.111.041302}{{\em Phys. Rev. Lett.} {\bfseries 111} (2013) 041302}, \href{https://arxiv.org/abs/1304.3461}{{\ttfamily arXiv:1304.3461 [hep-ph]}}.

\bibitem{Hochberg:2016sqx}
Y.~Hochberg, T.~Lin, and K.~M. Zurek, ``{Absorption of light dark matter in semiconductors},'' \href{https://dx.doi.org/10.1103/PhysRevD.95.023013}{{\em Phys. Rev. D} {\bfseries 95} no.~2, (2017) 023013}, \href{https://arxiv.org/abs/1608.01994}{{\ttfamily arXiv:1608.01994 [hep-ph]}}.

\bibitem{Bloch:2016sjj}
I.~M. Bloch, R.~Essig, K.~Tobioka, T.~Volansky, and T.-T. Yu, ``{Searching for Dark Absorption with Direct Detection Experiments},'' \href{https://dx.doi.org/10.1007/JHEP06(2017)087}{{\em JHEP} {\bfseries 06} (2017) 087}, \href{https://arxiv.org/abs/1608.02123}{{\ttfamily arXiv:1608.02123 [hep-ph]}}.

\bibitem{An:2020bxd}
H.~An, M.~Pospelov, J.~Pradler, and A.~Ritz, ``{New limits on dark photons from solar emission and keV scale dark matter},'' \href{https://dx.doi.org/10.1103/PhysRevD.102.115022}{{\em Phys. Rev. D} {\bfseries 102} (2020) 115022}, \href{https://arxiv.org/abs/2006.13929}{{\ttfamily arXiv:2006.13929 [hep-ph]}}.

\bibitem{SENSEI:2020dpa}
{\bfseries SENSEI} Collaboration, L.~Barak {\em et~al.}, ``{SENSEI: Direct-Detection Results on sub-GeV Dark Matter from a New Skipper-CCD},'' \href{https://dx.doi.org/10.1103/PhysRevLett.125.171802}{{\em Phys. Rev. Lett.} {\bfseries 125} no.~17, (2020) 171802}, \href{https://arxiv.org/abs/2004.11378}{{\ttfamily arXiv:2004.11378 [astro-ph.CO]}}.

\bibitem{XENON:2017lvq}
{\bfseries XENON} Collaboration, E.~Aprile {\em et~al.}, ``{The XENON1T Dark Matter Experiment},'' \href{https://dx.doi.org/10.1140/epjc/s10052-017-5326-3}{{\em Eur. Phys. J. C} {\bfseries 77} no.~12, (2017) 881}, \href{https://arxiv.org/abs/1708.07051}{{\ttfamily arXiv:1708.07051 [astro-ph.IM]}}.

\bibitem{Essig:2011nj}
R.~Essig, J.~Mardon, and T.~Volansky, ``{Direct Detection of Sub-GeV Dark Matter},'' \href{https://dx.doi.org/10.1103/PhysRevD.85.076007}{{\em Phys. Rev. D} {\bfseries 85} (2012) 076007}, \href{https://arxiv.org/abs/1108.5383}{{\ttfamily arXiv:1108.5383 [hep-ph]}}.

\bibitem{Arias:2012az}
P.~Arias, D.~Cadamuro, M.~Goodsell, J.~Jaeckel, J.~Redondo, and A.~Ringwald, ``{WISPy Cold Dark Matter},'' \href{https://dx.doi.org/10.1088/1475-7516/2012/06/013}{{\em JCAP} {\bfseries 06} (2012) 013}, \href{https://arxiv.org/abs/1201.5902}{{\ttfamily arXiv:1201.5902 [hep-ph]}}.

\bibitem{Fabbrichesi:2020wbt}
M.~Fabbrichesi, E.~Gabrielli, and G.~Lanfranchi, ``{The Dark Photon},'' \href{https://arxiv.org/abs/2005.01515}{{\ttfamily arXiv:2005.01515 [hep-ph]}}.

\bibitem{Redondo:2015iea}
J.~Redondo, ``{Atlas of solar hidden photon emission},'' \href{https://dx.doi.org/10.1088/1475-7516/2015/07/024}{{\em JCAP} {\bfseries 07} (2015) 024}, \href{https://arxiv.org/abs/1501.07292}{{\ttfamily arXiv:1501.07292 [hep-ph]}}.

\bibitem{An:2013yfc}
H.~An, M.~Pospelov, and J.~Pradler, ``{New stellar constraints on dark photons},'' \href{https://dx.doi.org/10.1016/j.physletb.2013.07.008}{{\em Phys. Lett. B} {\bfseries 725} (2013) 190--195}, \href{https://arxiv.org/abs/1302.3884}{{\ttfamily arXiv:1302.3884 [hep-ph]}}.

\bibitem{Iles:2024zka}
E.~Iles, S.~Heeba, and K.~Schutz, ``{Direct Detection of the Millicharged Background},'' \href{https://arxiv.org/abs/2407.21096}{{\ttfamily arXiv:2407.21096 [hep-ph]}}.

\bibitem{Stueckelberg:1938hvi}
E.~C.~G. Stueckelberg, ``{Interaction energy in electrodynamics and in the field theory of nuclear forces},'' \href{https://dx.doi.org/10.5169/seals-110852}{{\em Helv. Phys. Acta} {\bfseries 11} (1938) 225--244}.

\bibitem{Cline:2024wja}
J.~M. Cline and G.~Herrera, ``{Plausible constraints and inflationary production for dark photons},'' \href{https://arxiv.org/abs/2409.13818}{{\ttfamily arXiv:2409.13818 [hep-ph]}}.

\bibitem{Berlin:2023gvx}
A.~Berlin, R.~Tito~D'Agnolo, S.~A.~R. Ellis, and J.~I. Radkovski, ``{Signals of millicharged dark matter in light-shining-through-wall experiments},'' \href{https://dx.doi.org/10.1007/JHEP08(2023)017}{{\em JHEP} {\bfseries 08} (2023) 017}, \href{https://arxiv.org/abs/2305.05684}{{\ttfamily arXiv:2305.05684 [hep-ph]}}.

\bibitem{Berlin:2022hmt}
A.~Berlin, J.~A. Dror, X.~Gan, and J.~T. Ruderman, ``{Millicharged relics reveal massless dark photons},'' \href{https://dx.doi.org/10.1007/JHEP05(2023)046}{{\em JHEP} {\bfseries 05} (2023) 046}, \href{https://arxiv.org/abs/2211.05139}{{\ttfamily arXiv:2211.05139 [hep-ph]}}.

\bibitem{Redondo:2013lna}
J.~Redondo and G.~Raffelt, ``{Solar constraints on hidden photons re-visited},'' \href{https://dx.doi.org/10.1088/1475-7516/2013/08/034}{{\em JCAP} {\bfseries 08} (2013) 034}, \href{https://arxiv.org/abs/1305.2920}{{\ttfamily arXiv:1305.2920 [hep-ph]}}.

\bibitem{Redondo:2008aa}
J.~Redondo, ``{Helioscope Bounds on Hidden Sector Photons},'' \href{https://dx.doi.org/10.1088/1475-7516/2008/07/008}{{\em JCAP} {\bfseries 07} (2008) 008}, \href{https://arxiv.org/abs/0801.1527}{{\ttfamily arXiv:0801.1527 [hep-ph]}}.

\bibitem{OShea:2023gqn}
T.~O'Shea, M.~Giannotti, I.~G. Irastorza, L.~M. Plasencia, J.~Redondo, J.~Ruz, and J.~K. Vogel, ``{Prospects on the detection of solar dark photons by the International Axion Observatory},'' \href{https://dx.doi.org/10.1088/1475-7516/2024/06/070}{{\em JCAP} {\bfseries 06} (2024) 070}, \href{https://arxiv.org/abs/2312.10150}{{\ttfamily arXiv:2312.10150 [hep-ph]}}.

\bibitem{Ahlers:2007rd}
M.~Ahlers, H.~Gies, J.~Jaeckel, J.~Redondo, and A.~Ringwald, ``{Light from the hidden sector},'' \href{https://dx.doi.org/10.1103/PhysRevD.76.115005}{{\em Phys. Rev. D} {\bfseries 76} (2007) 115005}, \href{https://arxiv.org/abs/0706.2836}{{\ttfamily arXiv:0706.2836 [hep-ph]}}.

\bibitem{Graham:2014sha}
P.~W. Graham, J.~Mardon, S.~Rajendran, and Y.~Zhao, ``{Parametrically enhanced hidden photon search},'' \href{https://dx.doi.org/10.1103/PhysRevD.90.075017}{{\em Phys. Rev. D} {\bfseries 90} no.~7, (2014) 075017}, \href{https://arxiv.org/abs/1407.4806}{{\ttfamily arXiv:1407.4806 [hep-ph]}}.

\bibitem{An:2014twa}
H.~An, M.~Pospelov, J.~Pradler, and A.~Ritz, ``{Direct Detection Constraints on Dark Photon Dark Matter},'' \href{https://dx.doi.org/10.1016/j.physletb.2015.06.018}{{\em Phys. Lett. B} {\bfseries 747} (2015) 331--338}, \href{https://arxiv.org/abs/1412.8378}{{\ttfamily arXiv:1412.8378 [hep-ph]}}.

\bibitem{Scherer:2024uui}
H.~Sch\'erer and K.~Schutz, ``{Photon self-energy at all temperatures and densities in all of phase space},'' \href{https://arxiv.org/abs/2405.18466}{{\ttfamily arXiv:2405.18466 [hep-ph]}}.

\bibitem{Dressel_Grüner_2002}
M.~Dressel and G.~Grüner, {\em Electrodynamics of Solids: Optical Properties of Electrons in Matter}.
\newblock Cambridge University Press, 2002.

\bibitem{Raffelt:1996wa}
G.~G. Raffelt, {\em {Stars as laboratories for fundamental physics}}.
\newblock 5, 1996.

\bibitem{Gondolo:2008dd}
P.~Gondolo and G.~G. Raffelt, ``{Solar neutrino limit on axions and keV-mass bosons},'' \href{https://dx.doi.org/10.1103/PhysRevD.79.107301}{{\em Phys. Rev. D} {\bfseries 79} (2009) 107301}, \href{https://arxiv.org/abs/0807.2926}{{\ttfamily arXiv:0807.2926 [astro-ph]}}.

\bibitem{XENON:2019gfn}
{\bfseries XENON} Collaboration, E.~Aprile {\em et~al.}, ``{Light Dark Matter Search with Ionization Signals in XENON1T},'' \href{https://dx.doi.org/10.1103/PhysRevLett.123.251801}{{\em Phys. Rev. Lett.} {\bfseries 123} no.~25, (2019) 251801}, \href{https://arxiv.org/abs/1907.11485}{{\ttfamily arXiv:1907.11485 [hep-ex]}}.

\bibitem{XENON:2021qze}
{\bfseries XENON} Collaboration, E.~Aprile {\em et~al.}, ``{Emission of single and few electrons in XENON1T and limits on light dark matter},'' \href{https://dx.doi.org/10.1103/PhysRevD.106.022001}{{\em Phys. Rev. D} {\bfseries 106} no.~2, (2022) 022001}, \href{https://arxiv.org/abs/2112.12116}{{\ttfamily arXiv:2112.12116 [hep-ex]}}.

\bibitem{Betz:2013dza}
M.~Betz, F.~Caspers, M.~Gasior, M.~Thumm, and S.~W. Rieger, ``{First results of the CERN Resonant Weakly Interacting sub-eV Particle Search (CROWS)},'' \href{https://dx.doi.org/10.1103/PhysRevD.88.075014}{{\em Phys. Rev. D} {\bfseries 88} no.~7, (2013) 075014}, \href{https://arxiv.org/abs/1310.8098}{{\ttfamily arXiv:1310.8098 [physics.ins-det]}}.

\bibitem{Romanenko:2023irv}
A.~Romanenko {\em et~al.}, ``{Search for Dark Photons with Superconducting Radio Frequency Cavities},'' \href{https://dx.doi.org/10.1103/PhysRevLett.130.261801}{{\em Phys. Rev. Lett.} {\bfseries 130} no.~26, (2023) 261801}, \href{https://arxiv.org/abs/2301.11512}{{\ttfamily arXiv:2301.11512 [hep-ex]}}.

\bibitem{McDermott:2019lch}
S.~D. McDermott and S.~J. Witte, ``{Cosmological evolution of light dark photon dark matter},'' \href{https://dx.doi.org/10.1103/PhysRevD.101.063030}{{\em Phys. Rev. D} {\bfseries 101} no.~6, (2020) 063030}, \href{https://arxiv.org/abs/1911.05086}{{\ttfamily arXiv:1911.05086 [hep-ph]}}.

\bibitem{Caputo:2020bdy}
A.~Caputo, H.~Liu, S.~Mishra-Sharma, and J.~T. Ruderman, ``{Dark Photon Oscillations in Our Inhomogeneous Universe},'' \href{https://dx.doi.org/10.1103/PhysRevLett.125.221303}{{\em Phys. Rev. Lett.} {\bfseries 125} no.~22, (2020) 221303}, \href{https://arxiv.org/abs/2002.05165}{{\ttfamily arXiv:2002.05165 [astro-ph.CO]}}.

\bibitem{Chluba:2024wui}
J.~Chluba, B.~Cyr, and M.~C. Johnson, ``{Revisiting Dark Photon Constraints from CMB Spectral Distortions},'' \href{https://arxiv.org/abs/2409.12115}{{\ttfamily arXiv:2409.12115 [astro-ph.CO]}}.

\bibitem{Arsenadze:2024ywr}
G.~Arsenadze, A.~Caputo, X.~Gan, H.~Liu, and J.~T. Ruderman, ``{Shaping Dark Photon Spectral Distortions},'' \href{https://arxiv.org/abs/2409.12940}{{\ttfamily arXiv:2409.12940 [astro-ph.CO]}}.

\bibitem{Jaeckel:2010xx}
J.~Jaeckel and S.~Roy, ``{Spectroscopy as a test of Coulomb's law: A Probe of the hidden sector},'' \href{https://dx.doi.org/10.1103/PhysRevD.82.125020}{{\em Phys. Rev. D} {\bfseries 82} (2010) 125020}, \href{https://arxiv.org/abs/1008.3536}{{\ttfamily arXiv:1008.3536 [hep-ph]}}.

\bibitem{Kroff:2020zhp}
D.~Kroff and P.~C. Malta, ``{Constraining hidden photons via atomic force microscope measurements and the Plimpton-Lawton experiment},'' \href{https://dx.doi.org/10.1103/PhysRevD.102.095015}{{\em Phys. Rev. D} {\bfseries 102} no.~9, (2020) 095015}, \href{https://arxiv.org/abs/2008.02209}{{\ttfamily arXiv:2008.02209 [hep-ph]}}.

\bibitem{SENSEI:2023zdf}
{\bfseries SENSEI} Collaboration, P.~Adari {\em et~al.}, ``{SENSEI: First Direct-Detection Results on sub-GeV Dark Matter from SENSEI at SNOLAB},'' \href{https://arxiv.org/abs/2312.13342}{{\ttfamily arXiv:2312.13342 [astro-ph.CO]}}.

\bibitem{SENSEI:2024yyt}
{\bfseries SENSEI} Collaboration, I.~M. Bloch {\em et~al.}, ``{SENSEI at SNOLAB: Single-Electron Event Rate and Implications for Dark Matter},'' \href{https://arxiv.org/abs/2410.18716}{{\ttfamily arXiv:2410.18716 [astro-ph.CO]}}.

\bibitem{2022arXiv220210518A}
A.~{Aguilar-Arevalo}, F.~{Alcalde Bessia}, {\em et~al.}, ``{The Oscura Experiment},'' \href{https://dx.doi.org/10.48550/arXiv.2202.10518}{{\em arXiv e-prints} (Feb., 2022) arXiv:2202.10518}, \href{https://arxiv.org/abs/2202.10518}{{\ttfamily arXiv:2202.10518 [astro-ph.IM]}}.

\bibitem{Mizutani_2001}
U.~Mizutani, {\em Introduction to the Electron Theory of Metals}.
\newblock Cambridge University Press, 2001.

\bibitem{SuperCDMS:2016wui}
{\bfseries SuperCDMS} Collaboration, R.~Agnese {\em et~al.}, ``{Projected Sensitivity of the SuperCDMS SNOLAB experiment},'' \href{https://dx.doi.org/10.1103/PhysRevD.95.082002}{{\em Phys. Rev. D} {\bfseries 95} no.~8, (2017) 082002}, \href{https://arxiv.org/abs/1610.00006}{{\ttfamily arXiv:1610.00006 [physics.ins-det]}}.

\bibitem{Chen:2021tap}
{\bfseries Ricochet} Collaboration, R.~Chen, H.~D. Pinckney, E.~Figueroa-Feliciano, Z.~Hong, and B.~Schmidt, ``{Transition Edge Sensor Chip Design of a Modular CE$\nu$NS Detector for the Ricochet Experiment},'' \href{https://dx.doi.org/10.1007/s10909-022-02927-1}{{\em J. Low Temp. Phys.} {\bfseries 211} no.~5-6, (2023) 237--247}, \href{https://arxiv.org/abs/2111.05757}{{\ttfamily arXiv:2111.05757 [physics.ins-det]}}.

\bibitem{Ricochet:2023yek}
{\bfseries Ricochet} Collaboration, C.~Augier {\em et~al.}, ``{Results from a prototype TES detector for the Ricochet experiment},'' \href{https://dx.doi.org/10.1016/j.nima.2023.168765}{{\em Nucl. Instrum. Meth. A} {\bfseries 1057} (2023) 168765}, \href{https://arxiv.org/abs/2304.14926}{{\ttfamily arXiv:2304.14926 [physics.ins-det]}}.

\bibitem{Fuss:2022fxe}
P.~Adari {\em et~al.}, ``{EXCESS workshop: Descriptions of rising low-energy spectra},'' \href{https://dx.doi.org/10.21468/SciPostPhysProc.9.001}{{\em SciPost Phys. Proc.} {\bfseries 9} (2022) 001}, \href{https://arxiv.org/abs/2202.05097}{{\ttfamily arXiv:2202.05097 [astro-ph.IM]}}.

\bibitem{Schwarz:2015lqa}
M.~Schwarz, E.-A. Knabbe, A.~Lindner, J.~Redondo, A.~Ringwald, M.~Schneide, J.~Susol, and G.~Wiedemann, ``{Results from the Solar Hidden Photon Search (SHIPS)},'' \href{https://dx.doi.org/10.1088/1475-7516/2015/08/011}{{\em JCAP} {\bfseries 08} (2015) 011}, \href{https://arxiv.org/abs/1502.04490}{{\ttfamily arXiv:1502.04490 [hep-ph]}}.

\bibitem{harris2020array}
C.~R. Harris, K.~J. Millman, {\em et~al.}, ``Array programming with {NumPy},'' \href{https://dx.doi.org/10.1038/s41586-020-2649-2}{{\em Nature} {\bfseries 585} no.~7825, (Sept., 2020) 357--362}. \url{https://doi.org/10.1038/s41586-020-2649-2}.

\bibitem{2020SciPy-NMeth}
P.~Virtanen, R.~Gommers, {\em et~al.}, ``{{SciPy} 1.0: Fundamental Algorithms for Scientific Computing in Python},'' \href{https://dx.doi.org/10.1038/s41592-019-0686-2}{{\em Nature Methods} {\bfseries 17} (2020) 261--272}.

\bibitem{Hunter:2007}
J.~D. Hunter, ``Matplotlib: A 2d graphics environment,'' \href{https://dx.doi.org/10.1109/MCSE.2007.55}{{\em Computing in Science \& Engineering} {\bfseries 9} no.~3, (2007) 90--95}.

\bibitem{Braaten:1993jw}
E.~Braaten and D.~Segel, ``{Neutrino energy loss from the plasma process at all temperatures and densities},'' \href{https://dx.doi.org/10.1103/PhysRevD.48.1478}{{\em Phys. Rev. D} {\bfseries 48} (1993) 1478--1491}, \href{https://arxiv.org/abs/hep-ph/9302213}{{\ttfamily arXiv:hep-ph/9302213}}.

\bibitem{Vinyoles:2016djt}
N.~Vinyoles, A.~M. Serenelli, F.~L. Villante, S.~Basu, J.~Bergstr\"om, M.~C. Gonzalez-Garcia, M.~Maltoni, C.~Pe\~na Garay, and N.~Song, ``{A new Generation of Standard Solar Models},'' \href{https://dx.doi.org/10.3847/1538-4357/835/2/202}{{\em Astrophys. J.} {\bfseries 835} no.~2, (2017) 202}, \href{https://arxiv.org/abs/1611.09867}{{\ttfamily arXiv:1611.09867 [astro-ph.SR]}}.

\bibitem{Turck-Chieze:2001aug}
S.~Turck-Chieze {\em et~al.}, ``{Solar neutrino emission deduced from a seismic model},'' \href{https://dx.doi.org/10.1086/321726}{{\em Astrophys. J. Lett.} {\bfseries 555} (2001) L69--L73}.

\bibitem{Rodrigues:2020xpt}
D.~Rodrigues {\em et~al.}, ``{Absolute measurement of the Fano factor using a Skipper-CCD},'' \href{https://dx.doi.org/10.1016/j.nima.2021.165511}{{\em Nucl. Instrum. Meth. A} {\bfseries 1010} (2021) 165511}, \href{https://arxiv.org/abs/2004.11499}{{\ttfamily arXiv:2004.11499 [physics.ins-det]}}.

\end{thebibliography}\endgroup

\appendix

\section{In-medium massive photon dispersion relations}
\label{app:dispersion_relation}

The propagation of a massive gauge field in a bath of charged particles is modified by its interactions with the medium constituents.
The situation is similar to the case of massless gauge bosons, which is described in detail in Ref.~\cite{Raffelt:1996wa}.
However, the presence of a photon mass introduces relevant differences, the most important one being the existence of a longitudinal mode that can propagate in vacuum.

In the presence of a photon mass gauge freedom is lost, and charge conservation through the continuity equation requires using Lorentz gauge $\partial_\nu A^\nu=0$.
The inhomogeneous Maxwell's equations can be thus written as
\begin{equation}
    (\Box + m^2) A^\nu = J^\nu.
\end{equation}
Assuming a linear response, the effect of the medium can be modelled as an induced current $J^\mu_{\rm ind} = -\Pi^{\mu\nu} A_\nu$, where the polarization tensor $\Pi$ encodes the medium properties.
The Maxwell equations in Fourier space become
\begin{equation}
     \left[ -(k_\rho k^\rho + m^2)g^{\mu\nu} + \Pi^{\mu\nu} \right] A_\nu = J^\mu_{\rm ext},
\end{equation}
where $J_{\rm ext}$ denotes any external current.

In a parity-conserving medium, the polarization tensor can be decomposed as
\begin{equation}\label{eq:photon_self_energy}
    \Pi^{\mu\nu} = \pi_L P^{\mu\nu}_L + \pi_T  (P^{\mu\nu}_{T_1} + P^{\mu\nu}_{T_2}),
\end{equation}
where $P_L$ projects onto the longitudinal component and $P_{T_{1,2}}$ onto the two basis vectors spanning the transverse plane (e.g.~cartesian or circular polarization vectors).
The dispersion relations of longitudinal and transverse modes result from solving the homogeneous Maxwell equation for each mode, which reads
\begin{equation}
    -\omega^2 +k^2 + m^2 + \mathrm{Re}\,\pi_a(\omega, k) = 0,
\end{equation}
for $a = L,\, T_1,\, T_2$.
This yields an implicit equation for the frequency $\omega$ of a mode with a given wavenumber $k$.
The imaginary part of the polarization tensor describes photon absorption and production in the medium.
Another important quantity is the vertex renormalization constant arising from the photon wavefunction renormalization,
\begin{equation}
    Z_a^{-1} = 1 - \frac{\partial \pi_a}{\partial \omega^2}.
\end{equation}
This quantity effectively renormalizes the charge of the external particles interacting with the gauge boson.

In thermal field theory, $\Pi$ is the self-energy of the photon, and has been calculated to various orders in perturbation theory for a range of environments~\cite{Braaten:1993jw}.
The simplest and most relevant case for our purposes is that of a nonrelativistic, nondegenerate plasma, for which $\mathrm{Re}\,\pi_T = \omega_p^2$ to first order in QED.
Here, we have defined the plasma frequency of the fermionic charge carriers $\psi$,
\begin{equation}
    \omega_p^2 = \frac{4\pi\alpha n_\psi}{m_\psi},
\end{equation}
which only depends on their density $n_\psi$ and mass $m_\psi$ in the zero-temperature limit ($T\ll m_\psi$).
The dispersion relation of transverse modes is simply
\begin{equation}
    \omega^2 = k^2 + m^2 + \omega_p^2,
\end{equation}
which is that of a massive particle with effective mass $ m^2 + \omega_p^2$.
The corresponding renormalization factor is $Z_T=1$ to this order in QED.

\begin{figure*}[t]
\centering
\centerline{\includegraphics[width=0.5\linewidth]{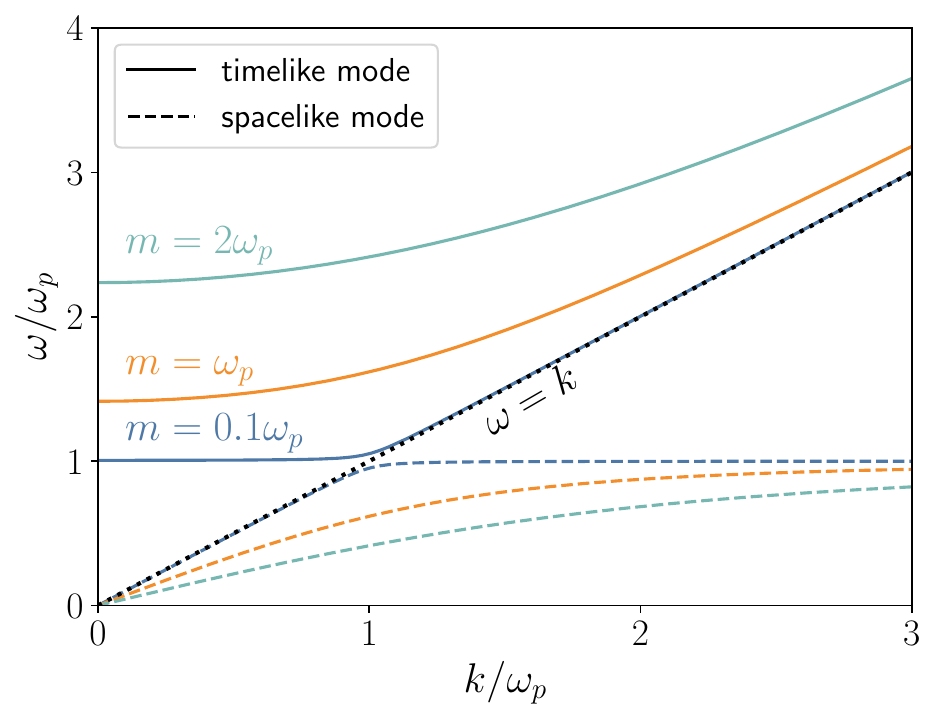}
\includegraphics[width=0.5\linewidth]{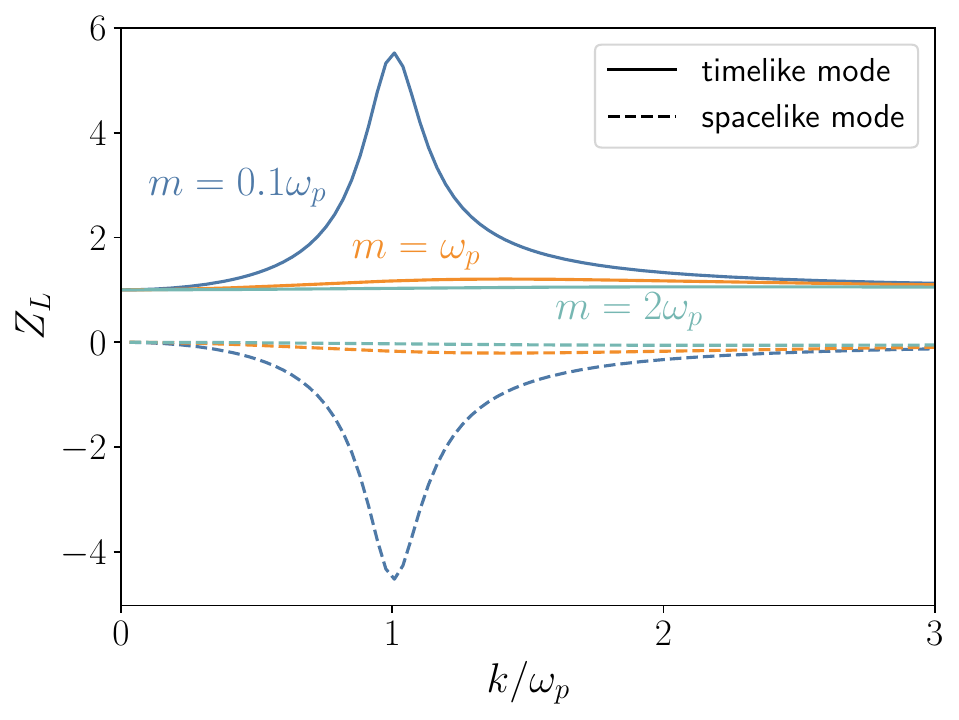}}
\caption{
Dispersion relation (left) and renormalization factor (right) for the longitudinal modes of a massive gauge boson propagating in a nonrelativistic nondegenerate plasma.
Different colors denote different values of the vacuum mass relative to the plasma frequency, while different line styles represent the two branches of the dispersion relation. The black dotted line on the left represents the massless photon dispersion relation in vacuum.
}
\label{fig:dispersion_relation}
\end{figure*}

The longitudinal component of the polarization tensor is related to the transverse one by
\begin{equation}
\label{eq:piL_piT}
    \pi_L = \frac{\omega^2 - k^2}{\omega^2} \pi_T
\end{equation}
Thus, the dispersion relation for longitudinal modes can be found by solving
\begin{equation}\label{eq:dispersion_relation_L}
    \omega^2 - k^2 = m^2 + \omega_p^2 \frac{\omega^2-k^2}{\omega^2},
\end{equation}
and the renormalization factor is in this case
\begin{equation}
    Z_L = \frac{\omega^4}{\omega^4 - k^2\omega_p^2}.
\end{equation}
If $m=0$, Eq.~\eqref{eq:dispersion_relation_L} is solved by $\omega = \omega_p$ independently of wavenumber (i.e.~plasmons oscillate at their plasma frequency regardless of their momentum), and thus $Z_L=\omega_p^2/(\omega_p^2-k^2)$.
The solutions for $m\neq 0$ are shown in Fig.~\ref{fig:dispersion_relation} as a function of $k$ for different values of $m$.
The dispersion relation of a massive gauge field in a plasma develops two branches, one being time-like and the other one space-like.
When $m\simeq\omega_p$, the time-like mode behaves like a plasmon with $\omega\simeq\omega_p$ for small wavenumber, while at large $k\gtrsim\omega_p$ the dispersion relation approaches the vacuum one $\omega^2 = k^2+m^2$. 
As the plasma becomes more dilute and $\omega_p$ becomes much smaller than $m$, the time-like mode approaches the vacuum dispersion relation while the space-like mode effectively decouples as $Z_L\rightarrow 0$ (this mode has no support outside of the plasma).
In the opposite $m\ll\omega_p$ limit, there is an avoided crossing between the asymptotic plasmon with $\omega=\omega_p$ for all $k$ and an almost massless mode corresponding to the vacuum longitudinal mode.

\section{Longitudinal massive photon emission from a sun-like star}\label{app:sun_L_emission}

Building on the formalism in Appendix~\ref{app:dispersion_relation}, here we calculate what the spectral emission of the sun would be if the photon were to suddenly acquire a small mass (so as not to directly affect atomic or other relevant processes).
For that, we take the solar model developed in Ref.~\cite{Vinyoles:2016djt}\footnote{We use the updated values publicly available at~\url{https://zenodo.org/records/10822316} for the SF3-GS98 model.} supplemented with the finer surface temperature and density values of Ref.~\cite{Turck-Chieze:2001aug}.
Near the surface, the precise values of the free electron density, as well as of the different hydrogen isotopes ($H^+$, $H^0$, and $H^-$), are calculated following~\cite{Redondo:2015iea}.

The crucial quantity to calculate is the self energy of transverse and longitudinal modes, which are related via Eq.~\eqref{eq:piL_piT}.
Their real and imaginary parts control the photon dispersion and absorption/production rates respectively, so one can write them as
\begin{align}
    \mathrm{Re}(\pi_T) &= m_T^2 = \frac{\omega^2}{\omega^2 - k^2} m_L^2 \,, \\
    \frac{-1}{\omega}\mathrm{Im}(\pi_L) &= \Gamma_T = \frac{\omega^2}{\omega^2 - k^2} \Gamma_L.
\end{align}
We can see that both the effective photon mass well as the damping rate are suppressed for the longitudinal modes if $\omega^2-k^2\ll\omega^2$.
With this, the equilibrium absorption and production rates are given by
\begin{align}
    \Gamma^{\rm abs}_{T/L} &= \left( 1-\me^{-\omega/T} \right)^{-1} \Gamma_{T/L},\\
    \frac{\diff\Gamma^{\rm prod}_{T/L}}{\diff\omega\,\diff V} &= \frac{\omega k}{2\pi^2} \frac{1}{ \me^{-\omega/T}-1} \Gamma_{T/L}.
\end{align}
In the sun, the photon self-energy gets contribution from plasma (Thomson scattering), free-free (bremsstrahlung), bound-free (photoelectric), and bound-bound (atomic transition) processes and can thus be written as~\cite{Redondo:2015iea},
\begin{equation}
    \pi_T = \pi_{\rm plas} + \pi_{\rm ff} + \pi_{\rm bf} + \pi_{\rm bb}.
\end{equation}
We calculate the contributions to the self-energy as a function of frequency and position in the sun following~\cite{Redondo:2015iea}.
At the frequencies of interest, the plasma term dominates the real part in the solar interior, while the other terms become relevant near the photosphere.
For the imaginary part, free-free emission and absorption is most relevant except near the surface, where all contributions must be taken into account.

With this, the luminosity of each polarization mode can be calculated as
\begin{equation}
    \frac{\diff L_{T/L}}{\diff\omega} = \int_{0}^{R_\odot} \mathop{\diff r} r^2 \, \frac{2\omega k}{\pi} \, \left( 1-\me^{-\omega/T} \right)^{-1} \Gamma_{T/L}\,P_{T/L}^{\rm esc},
\end{equation}
where the escape probability is
\begin{equation}
    P_{T/L}^{\rm esc}(r) = \mathrm{exp}\left( -\int_{r}^{R_\odot} \mathop{\diff\rho} \Gamma_{T/L}^{\rm abs}(\rho) \right).
\end{equation}
Here, $R_\odot$ is understood to be the radius of the sun at which the production rate is already negligible.
Note that one must be careful in distinguishing the vacuum wavenumber from the one inside the sun, which is modified due to the in-medium dispersion relation and is different for each polarization.

\begin{figure}[t]
\centering
\centerline{\includegraphics[width=0.95\linewidth]{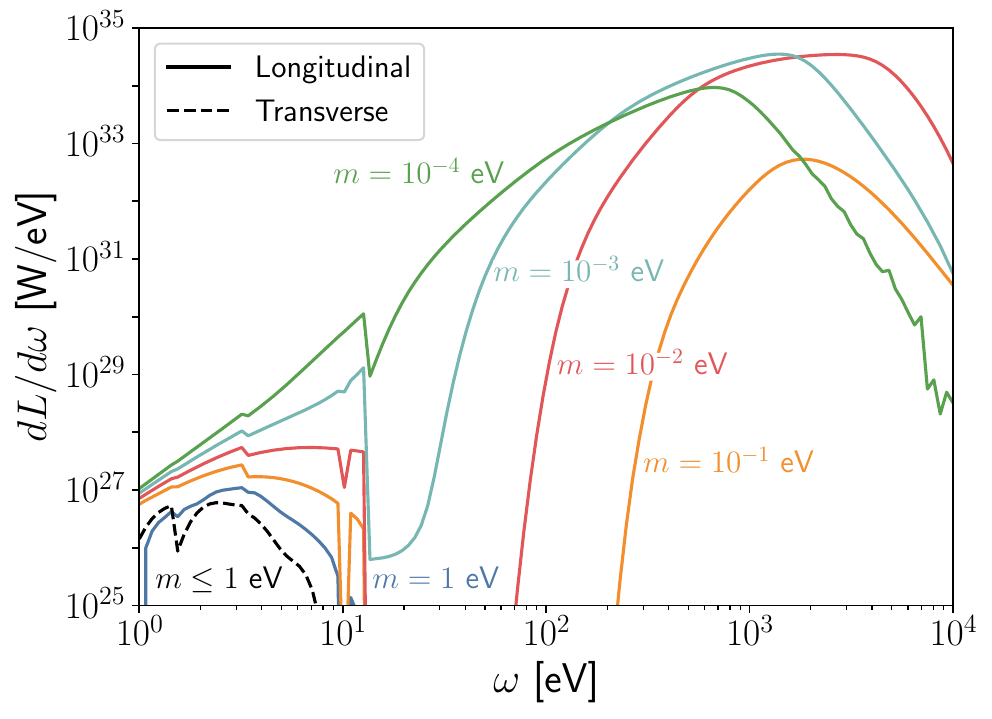}}
\caption{
Spectral luminosity of longitudinally (solid lines) and transverse (dashed line) polarized light from a sun-like star.
Different colors denote different values of the photon mass (the transverse emission is insensitive to it for the small masses shown).
The low frequency peak corresponds to surface emission near the photosphere, while the high frequency one is due to volume emission from the interior of the star.
}
\label{fig:L_emission_sun}
\end{figure}

The resulting luminosities are shown in Fig.~\ref{fig:L_emission_sun} for different values of the dark photon mass.
A small photon mass does not significantly affect the transverse mode emission, which still occurs at the solar photosphere since the solar interior is opaque to transverse photons.
The spectrum is peaked at visible frequencies and the total integrated luminosity is $\mathcal{O}(10^{26})$ watt.

However, the existence of a longitudinal polarization that can propagate out of the sun has dramatic consequences.
For sub-eV masses, the longitudinal mode is sufficiently weakly coupled to escape the interior of the sun, leading to a volumetric emission that greatly enhances the luminosity.
The spectrum develops a second, more pronounced peak at energies comparable to the core temperature of the sun, and the total luminosity can be enhanced by close to 10 orders of magnitude for dark photon masses in the meV range.
Clearly, such enormous luminosity would have a significant impact on the stellar structure.
Accounting for backreaction in the solar model is beyond the scope of this work, but we expect the system to shrink and heat up, possibly leading to hotter, more luminous, shorter-live stars.
We leave further studies of the impact of a dark photon mass on dark stellar evolution for future work.

\section{SENSEI limits}\label{app:SENSEI}

The SENSEI experiment is sensitive to absorption of dark photons depositing enough energy to excite electrons to the conduction band of silicon.
In Refs.~\cite{SENSEI:2023zdf,SENSEI:2024yyt}, model-independent limits are presented on the rate of $n\,e^-$ excitations, for $1\leq n\leq 10$, which we use to place a limit on the dark photon flux at earth.

The rate of absorption of massless dark photons in silicon can be calculated using Eq.~\eqref{eq:event_rate_volume} with $\rho=2.3\mathrm{g}/\mathrm{cm}^3$.
To calculate the volumetric absorption rate in Eq.~\eqref{eq:absorption_rate_volume}, we use the room-temperature values for the conductivity of silicon from~\cite{Hochberg:2016sqx}.
The deposited energy necessary to excite $n$ electrons in silicon is~\cite{SENSEI:2020dpa,Rodrigues:2020xpt}
\begin{equation}
    E_{ne^-} = 1.2\,\mathrm{eV} + (n-1)\,3.8\,\mathrm{eV}.
\end{equation}
Thus, the expected rate for $n$ electronic excitations is given by
\begin{equation}
    R_{ne^-} = \int_{E_{ne^-}}^{E_{(n+1)e^-}} \mathop{\diff\omega} \frac{\mathop{\diff R_{\rm vol}}}{\mathop{\mathrm{d}\omega}},
\end{equation}
The limits on $R_{ne^-}$ from Ref.~\cite{SENSEI:2023zdf} can thus be used to constrain $\epsilon^2$ times the dark photon flux.
We combine the the $4-10$ electron bins in which no events are observed and take the most constraining bin for each parameter point.

\section{XENON 1T limits}\label{app:XENON1T}

The S2 (secondary scintillation signal) event rate at the XENON1T experiment is calculated following Section~V of Ref.~\cite{XENON:2021qze}.
We assume that all electrons have a drift probability of $80\%$ and do not consider the impact of individually varying the depth at which the ionization electron is created.

To recast the single-electron analysis of~\cite{XENON:2021qze}, we employ a gain factor of 28.8 photoelectrons per primary electron and the selection efficiency shown in Fig.~10 of that work.
The calculated expected event rate as a function of S2 is compared with the limits shown in Fig.~11 of that paper to obtain constraints on the corresponding signal (we use the most constraining bin).

For the search presented in~\cite{XENON:2019gfn}, we use a gain factor of 33 and extract the selection efficiency from Fig.~2 in that work.
As in the previous case, the S2-distribution of events is compared to the observed one shown in Fig.~4 of~\cite{XENON:2019gfn}. To derive limits, we only keep the most constraining bin.

\end{document}